\documentclass{IEEEtran}
\usepackage{cite}
\usepackage{amsmath,amssymb,amsfonts,amsthm}
\usepackage{algpseudocode}
\usepackage[hyphens]{url}
\usepackage{algorithm}
\usepackage{algorithmicx}
\usepackage{graphicx}
\usepackage{float}
\usepackage{textcomp}
\usepackage{xcolor}
\usepackage{url}

\theoremstyle{definition}
\newtheorem{definition}{Definition}[section]

\theoremstyle{remark}

\usepackage{subfigure}

\title{Secure Access Control for DAG-based Distributed Ledgers}

\author{L.~Zhao\thanks{Lianna Zhao, Andrew Cullen, Pietro Ferraro and Robert Shorten are with the Dyson School of Design Engineering at Imperial College London.}, L.~Vigneri, A.~Cullen, W.~Sanders \thanks{William Sanders and Luigi Vigneri are with the IOTA Foundation.}, P.~Ferraro, and R.~Shorten}

\begin{document}

\maketitle

\begin{abstract}

Access control is a fundamental component of the design of distributed ledgers, influencing many aspects of their design, such as fairness, efficiency, traditional notions of network security, and adversarial attacks such as Denial-of-Service (DoS) attacks\footnote{Attackers attempt to put stress on the network by sending a large amount of transactions to other nodes.}.  In this work, we consider the security of a recently proposed access control protocol for Directed Acyclic Graph-based distributed ledgers. We present a number of attack scenarios and potential vulnerabilities of the protocol and introduce a number of additional features which enhance its resilience.  Specifically, a blacklisting algorithm, which is based on a reputation-weighted threshold, is introduced to handle both spamming and multi-rate malicious attackers. The introduction of a solidification request component is also introduced to ensure the fairness and consistency of network in the presence of attacks. Finally, a timestamp component is also introduced to maintain the consistency of the network in the presence of multi-rate attackers. Simulations to illustrate the efficacy and robustness  of the revised protocol are also described. 
\end{abstract}

\section{Introduction}
Over the last decade, Distributed Ledger Technologies (DLTs) have become a very popular topic both in industry and academic communities because of their broad applications across many vertical domains, such as, supply chain, smart cities~\cite{8496756} and decentralized finance~\cite{singh2019distributed}. In essence, a distributed ledger, just as the name suggests, is an immutable database shared across multiple agents in a decentralized manner, where participants have consensus on the contents of the database. DLT's are viewed as a game changing technology in many industries as they enable an immutable recording of exchange of assets, and can also be used to design cyberphysical systems \cite{ferraro2018distributed}.\newline

The best known DLT architectures is Blockchain~\cite{8496756}. While the inception of Blockchain represented a leap forward in achieving consensus and immutability in a Peer-to-Peer (P2P) network, its technological and design limitations make it unsuitable for many Internet-of-Things (IoT) applications \cite{dai2019blockchain}. For example, from a structural perspective, the incentive mechanism in Blockchain encourages the validation of large transactions over small ones, and tends to centralise power in the hands of a few powerful mining pools \cite{pilkington2016blockchain}. Moreover, the time interval between the creation of new blocks leads to low throughput and scalability issues that seriously limits the range of domains to which Blockchain can be applied.\newline

Recently, many researchers have suggested alternatives to the basic Blockchain structure. One such alternative is Direct-Acyclic-Graph (DAG) based DLTs~\cite{cullen2020resilience,li2020direct}. A typical DAG architecture is represented by the IOTA Tangle~\cite{popov2018tangle}. Instead of using a chain, every new incoming transaction can freely reference existing transactions in a graph structure. This means many transactions are verified in a parallel fashion, thereby realising a more scalable and high-throughput architecture. Unlike in the Blockchain where Proof-of-Work (PoW) is used to reach consensus among nodes, the IOTA Tangle employs a lightweight reputation-based voting mechanism to keep consistent ledger states, and PoW is only employed to make Denial-of-Service (DoS) attacks expensive~\cite{bentov2016cryptocurrencies}. Consequently, the PoW in the Tangle tends to be computationally much less expensive than the one used in Blockchain \cite{ferraro2018distributed}.\newline 

In this paper, we consider certain specific DAG-based ledgers that have been proposed for IoT applications. While their design overcomes some of the scalability and economic issues in traditional Blockchains, these DLTs also give rise to new challenges in terms of the design of access control mechanisms to support their operation. In fact, Blockchains have an intrinsic {\em filter} provided by the work performed by the miners to select which transactions should be added to the ledger. Removing PoW makes DAG-based DLTs more IoT-friendly, but also necessitates an explicit access control mechanism to guarantee fair writing rights to the network nodes. While the literature on congestion and access control for conventional networking applications is rich, a distinguishing feature of DLTs is that they must be designed to operate in a reliable manner in adversarial environments, and this mandates new line of research in the area of congestion and access control.\newline

Our objective here is to address this issue. We focus on the IOTA Tangle and its recently proposed reputation-based\footnote{In this paper, we consider reputation as a numeric value associated to a node depending on some characteristics. In principle, reputation should be difficult to gain and easy to lose; reputation may be simply computed according to the stake handled by nodes or by analysing node's behavior from a more complex point of view. However, the actual way of computing reputation is out of the scope of the paper. We also assume that nodes have perfect knowledge of reputation of all the other nodes. While this assumption may not be realistic, our access control algorithm can successfully deal with small inconsistencies in the reputation perception among different nodes.} access control algorithm \cite{cullenaccess}. However, our analysis can be applied to any distributed network that seeks to guarantee fair access and consistent databases across the network participants in an adversarial environment. The basic idea is to allow users with a larger amount of reputation\footnote{The measure of reputation in IOTA is called mana.} to have a larger impact on the system. Note that in DLTs, access to the ledger is highly competitive, affecting both consensus and the ability of users to generate revenue. Consequently, access control mechanism plays a very important role in DLTs. This paper studies the reference mechanism adopted as a part of the IOTA protocol, describes its vulnerabilities and proposes a number of additional countermeasures to fix these problems to be resistant to specific attack scenarios. In particular, we consider the following issues concerning the aforementioned protocol.\newline
\begin{itemize}
\item We analyse the security of~\cite{cullenaccess} against more advanced attacks than considered in the original work, e.g., nodes send different streams of transactions to different neighbours to harm consistency.\newline
\item We introduce several components to ensure ledger consistency across network participants (\textcolor{black} {see Definition IV.6}). More specifically,  we propose specific buffer management techniques, transaction reordering and improved gossip strategies to improve the network solidification process. (\textcolor{black} {For a discussion of {\em solidification} requests and timestamp ordering, please refer to V.B}).\newline
\item We propose an effective method to penalise attackers through blacklisting specific malicious flows (blacklisting strategy - please refer to IV.A).\newline
\item We evaluate and benchmark the robustness of the newly designed IOTA access control algorithm via extensive simulations.\newline
\end{itemize}

The remainder of this paper is organised as follows. In Section II, we give a brief description of access control mechanism for DLTs and some related network concepts. In Section III, we present Blockchain structure including its commonly used consensus mechanism and  DAG-based DLTs. In Section IV, a node model for our access control mechanism and definitions needed in this paper are provided. In Section V, we review the access control protocol described in~\cite{cullenaccess}, highlight vulnerabilities, and suggest modifications to suppress these vulnerabilities. Section VI presents simulations to illustrate the efficacy of our approach. Finally, we conclude our paper in Section VII.

\section{Comments on prior work}
The topic of designing access control algorithms, and more generally access control mechanisms, is a very rich one in the domain of the computer networking. That said, this topic, is relatively new when considering distributed ledgers, and the design of such algorithms in this specific context gives rise to new and original challenges due to the adversarial nature of the environments in which ledgers are designed to operate. This richness, and simultaneous sparsity, makes a compact discussion of prior work challenging. Our objective here is thus to highlight specific concepts that underpin the work described in this paper, and to provide a hint of the broader context. In terms of DLTs, recently several architectures, have been proposed incorporating access control modules both to guarantee fair writing rights to the network nodes \cite{maesa2017blockchain, outchakoucht2017dynamic}, and to prevent spamming attacks \cite{hu2020securing, he2018consensus}. These include both NANO and IOTA \cite{lemahieu2018nano, pinjala2019dcaci}. The NANO DLT is particularly interesting when highlighting the need for access control mechanisms. Specifically, the NANO ledger was subjected to attacks that could have been prevented by incorporating an appropriately designed access control mechanism\footnote{https://senatusspqr.medium.com/nanos-latest-innovation-feeless-spam- resistance-f16130b13598}. Indeed, the motivation for this present paper is the proposed IOTA access control algorithm \cite{cullenaccess} and the need to make this module robust against certain types of adversarial attacks. In terms of connection to traditional networking work, our work builds heavily on some of the most mature work in this area, with the caveat that algorithm design is revisited from the perspective of operation in adversarial environments, That said, much of the work reported in this paper builds on traditional scheduling algorithms such as {\em Defecit Round Robin} (DRR), and flow based access control, such as the {\em Additive-Increase Multiplicative-Decrease} (AIMD) algorithm \cite{corless2016aimd}. Such algorithms are well documented in many publications, and we do not repeat this discussion here. Nevertheless, the interested reader is referred to the following publications for more background on these and related topics \cite{lakshman1997performance} \cite{1995Analysis}\cite{hasegawa2001analysis} \cite{pan2003approximate} \cite{floyd1993random}. 

\section{Preliminaries}

\subsection{Blockchain}

\begin{figure}[ht]
\centering
\includegraphics[width=\columnwidth]{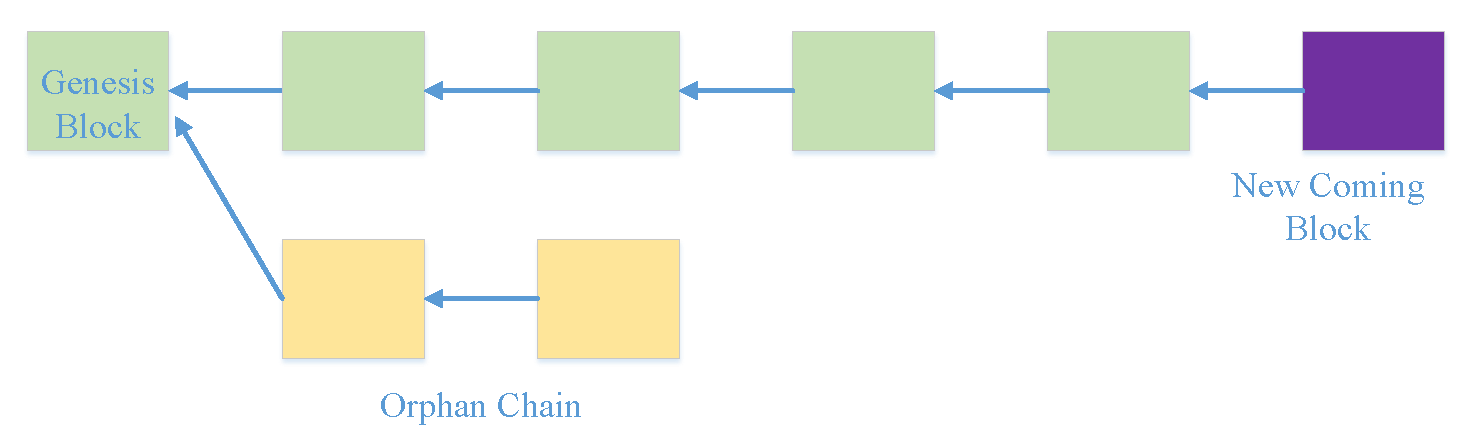}
\caption{A simplified Blockchain structure. The purple block is the new incoming block which need be added  to the chain. The green blocks (except the first block also named genesis block) are already added to Blockchain, yellow blocks are similar to green blocks, but in orphan chain.}
\label{fig: longestrule}
\end{figure}

As we have mentioned, Blockchain is the most famous DLT architecture and was introduced by Satoshi Nakamoto in the Bitcoin whitepaper~\cite{nakamoto2019bitcoin} in 2009. 
In Blockchain, as in all DLTs, data are replicated across all nodes in a P2P network without a third-party or central authority. Transparency and consistency are ensured as all nodes in the network maintain a copy of the ledger and independently verify each new record. This is done in a manner which makes it very difficult for any node to change the content of the ledger. This property is known as immutability. A high-level diagram of the Blockchain structure is depicted in Figure \ref{fig: longestrule} which we describe next.\newline

In Blockchain, special users called miners are responsible for gathering data and assembling them into blocks that are later added to the Blockchain. Immutability is then guaranteed by the fact that each block is connected to the previous one through a cryptographic mechanism that exploits PoW and hash functions~\cite{gervais2016security}. The aim of this hash-based PoW is to find a nonce value which is smaller than the current target value. A new block will be issued successfully by miners who successfully find such nonce.
This solves the problem of double-spending and Sybil attacks\footnote{Sybil attack means attackers are trying to get multiple identities in order to gain advantage in a reputation system\cite{douceur2002sybil}.} at the cost of a large computational expense and a low throughput. For more details, the interested reader can refer to~\cite{butun2020review}~\cite{cullen2021access}~\cite{xu2019blockchain}.\newline

An alternative to PoW is Proof-of-Stake (PoS)~\cite{yaga2019blockchain}. In PoS, instead of requiring miners to do computational work, the probability of successfully issuing blocks relies on a quantity called stake that represents a share of the total currency. Namely, the more stake a node has, the higher chance this node can be selected to issue a block. A variant of PoS is called Delegated Proof-of-Stake (DPoS) and is employed, for instance, in BitShares~\cite{schuh2017bitshares}, EOS~\cite{elrom2019eos} and Cosmos~\cite{kwon2019cosmos}. In DPoS, not all nodes are allowed to issue blocks. Every node holding stake votes for its trusted witnesses whose job is to issue and validate blocks as the representative of all nodes. Because there are fewer nodes that participate in block issuing and validating work, the issuing and confirmation speed of blocks results accelerated.

\subsection{DAG-based DLT}
\begin{figure}[ht]
	\centering
	\includegraphics[width=\columnwidth]{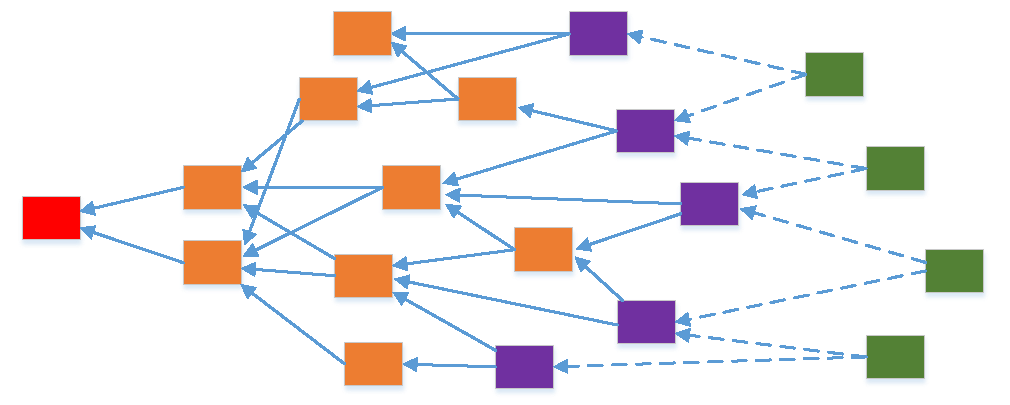}
	\caption{A simplified architecture of ledger }
	\label{fig: Tipselection}
\end{figure}

Recently a number of DLTs based in directed acyclic graphs (DAGs) have been developed, specifically with a focus on serving the needs of the IoT industry \cite{pervez2018comparative}. One such architecture is the IOTA Tangle. A DAG is a graph with no directed cycles and in the IOTA Tangle, each node in the graph represents a transaction. Roughly speaking, the IOTA Tangle operates as follows. New transactions arrive and are appended to the DAG in return for approving existing transactions that have been already added to the DAG but which have not yet been approved by other transaction. This is depicted in Figure \ref{fig: Tipselection}. Newly arriving transactions (green) randomly select unapproved transactions (purple) and check that these transactions are consistent with each other and the contents of the ledger. Once these selected purple transactions have been validated, they become orange, and the green validating transactions become purple. In the language of the IOTA Tangle, with reference to Figure \ref{fig: Tipselection}, if there is a directed path between transaction $i$ and transaction $j$, we say transaction $i$ directly/indirectly approves\footnote{https://blog.iota.org/the-tangle-an-illustrated-introduction-1618d3e140ad/} transaction $j$. When new transactions arrive, they validate up to eight (with two as a default) existing transactions which are chosen at random~\cite{popov2020coordicide}. 
For further details on the consensus protocol of the Tangle, the interested reader can refer to~\cite{popov2020coordicide}. \newline

\begin{figure}[ht]
	\centering
	\includegraphics[width=\columnwidth]{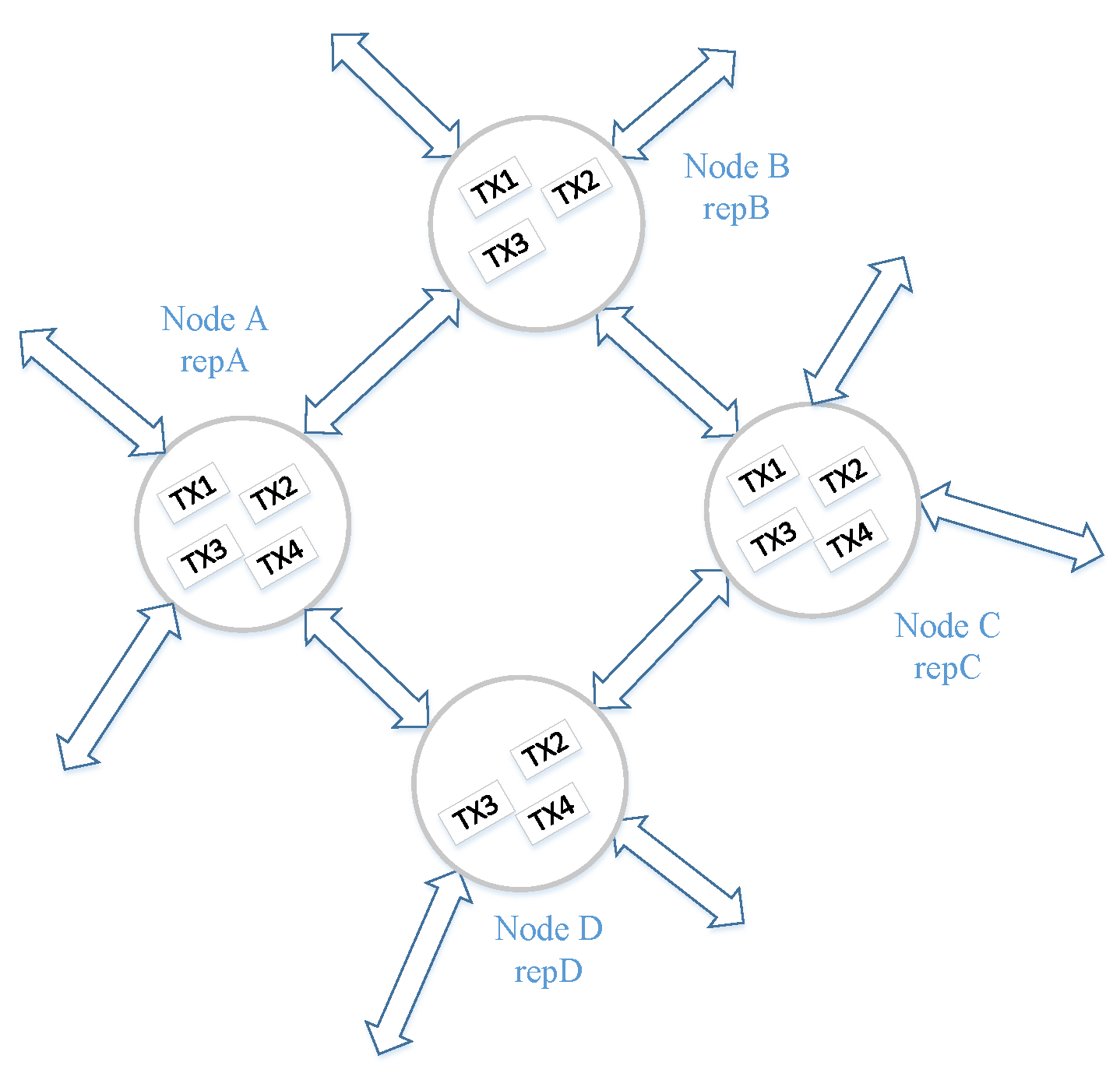}
	\caption{A peer-to-peer network}
	\label{fig: Network}
\end{figure}

While Figure \ref{fig: Tipselection} depicts the architecture of a ledger stored in a single device, multiple copies of the Tangle are stored and synchronized across many devices. This scenario is depicted in Figure \ref{fig: Network} and makes evident the need for an access control mechanism which in turn dictates the fairness properties of the ledger and allows the Tangle to resist spamming attacks that might cause the loss of synchronization of the different copies of the ledger in the network. Note that while access is controlled automatically in Blockchain through leader election (e.g., PoW, PoS), no such system naturally arises in the Tangle. Thus, the design of an access control mechanism for lightweight IoT-friendly DLTs, such as the IOTA Tangle, is particularly pressing.

\section{System model}\label{sec: model}

We consider a P2P network with a set of nodes $\mathcal{N}$ where $|\mathcal{N}| = N$. Each node is connected with $n$ neighbours where $n\ll N$. Node $m$ has a reputation value $rep_m\in\mathbb{R}^+$. Nodes perform two main tasks.\newline
\begin{itemize}
	\item Nodes verify the validity of existing transactions (e.g., verifying that the transaction has a valid signature) and participate in conflict (e.g., double spending) resolution. Discussion related to transaction validation and consensus are outside the scope of the paper, but can be found in~\cite{he2021consensus}~\cite{xiao2020survey}.\newline
	\item Nodes issue transactions to transfer data or value transactions. Each transaction must include, among other information, the ID of the issuer, the list of transactions which approves, a timestamp indicating the local current issuing time.\newline
\end{itemize}

In line with~\cite{cullenaccess}, we describe here the main components of the protocol (more information can be found later in the section).\newline
\begin{itemize}
	\item Transactions received from neighbours are first filtered to remove invalid ones according to some rules: some typical filters include signature validation and removal of duplicates and old transactions.\newline
	\item Filtered transactions are added to the inbox of the node; specifically, the inbox is split into $N$ queues in order to differentiate the node issuers, where $q_{m_i}$ denotes transactions issued by node $i$ in node $m$'s inbox.\newline
	\item We mentioned that nodes can also issue transactions themselves. The issuing rate of each node is controlled by a rate setting algorithm according to the node's reputation and its inbox length.\newline
	\item A scheduling mechanism is employed to schedule transactions from the inbox. The protocol sets a fixed global transaction writing power $\nu$,  where $\nu$ is the rate at which this writing work is done. Hence, the dissemination rate, $DR$ (see Definition IV.4), can be at most $\nu$. The reader should note that fixing the scheduling rate is a fundamental component for the successful operation of the protocol defined in \cite{cullenaccess}. We denote that the rate at which node $m$ issues transactions as  $\lambda_m$. In the current version of the protocol this minimum rate is set as  $\lambda_m \leq \frac{\nu \cdot rep_m}{\sum_{i \in \mathcal{N}}{rep_i}}$.\newline
	\item Scheduled transactions are then forwarded to neighbours through a gossip algorithm and added to the local version of the ledger (if they satisfy the consensus protocol of the DLT).\newline
\end{itemize}

Depending on the transaction issuing rate, a node can be said to be in one of four possible states.\newline
\begin{itemize}  
	\item \emph{Inactive node}: A node is said to be inactive if it is not issuing transactions -- it only stores the ledger updates and participates in conflict resolution.\newline
	\item \emph{Content node}: We model the issuing rate of a content node $m$ as a Poisson process with a rate parameter $\lambda_m \leq \tilde{\lambda}_m = \frac{\nu \cdot rep_m}{\sum_{i \in \mathcal{N}}{rep_i}}$, \textcolor{black}{ where, as we have mentioned, $\frac{\nu \cdot rep_m}{\sum_{i \in \mathcal{N}}{rep_i}}$ is the minimum allowed rate and $rep_m$ is the reputation of node $m$.} In other words, content nodes never exceed their fair proportion of the global writing power according to their reputation.\newline
	\item  \emph{Best-effort node}: A node is said to be best effort if it is issuing at rate $\lambda_m > \frac{\nu \cdot rep_m}{\sum_{i \in \mathcal{N}}{rep_i}}$, while obeying the restrictions imposed by the access control algorithm. In practice, if many nodes are inactive or issue transactions occasionally, which is likely to be the case, some nodes may want to exploit the unused bandwidth to issue more transactions than their content rate.\newline
	\item \emph{Malicious node}: A node is said to be malicious if it does not follow the rules imposed by the access control algorithm. Such nodes try to harm the network and affect consistency and fairness by introducing congestion and degrading network performance.\newline
\end{itemize}

To help the exposition, the notation used in the remainder of the paper is summarized in Tables~\ref{tab: notation1} and~\ref{tab: notation2}.
\begin{table}[ht]
	\caption{Notation for node and network model.}
	\centering
	\begin{tabular}{c|l}
		\hline\hline
		$rep_i$            & reputation of node $i$ \\
		$\mathcal{N}$      &  the set of all nodes\\
		$N$                & the total number of nodes\\
		$q_{m_{i}}$        &  the queue length for node $i$'s transactions in node $m$'s inbox\\
		$\lambda_m$        & issuing rate of node $m$\\
		$\nu$              & global transaction writing power \\
		$DR_i$ 	           & the dissemination rate of node $i$\\
		$DR$     	       & the dissemination rate of all transactions\\
		$DC_{max}$     	       & the maximum deficit for each empty queue\\
		\hline\hline
	\end{tabular}
	\label{tab: notation1}
\end{table}

\begin{table}[ht]
	\caption{Notation for pseudocode.}
	\centering
	\begin{tabular}{c|l}
		\hline\hline
		$p$                    & transaction's parent \\
		$\epsilon$        & a given fixed time threshold \\
		$\sigma$          & given constant\\
		$ReqSolid$             & solidification requests list \\
		$Tran$                 & node's inbox transactions list \\
		$Scheduled$            & transactions list we have already scheduled \\
		$T_{BL_{i}}$            & the time when any honest node blacklist malicious nodes \\
		$\hat{q}_{m_{i}} $         & reputation-scaled inbox length for node $i$ \\
		$W_{BL}$                 & the threshold of blacklisting a node \\
		$T_{H_{i}}$            & the time interval between current time and $T_{BL_{i}}$ \\
		\hline\hline
	\end{tabular}
	\label{tab: notation2}
\end{table}
Finally, we now add a number of definitions to assist the exposition of the sequel.\newline

\begin{definition}[{\em Past cone}]\label{def: past cone}
The past cone of transactions $A$ is the set of all transactions which transaction $A$ approves either directly or indirectly.\newline
\end{definition}

\begin{figure}[ht]
	\centering
	\includegraphics[width=\columnwidth]{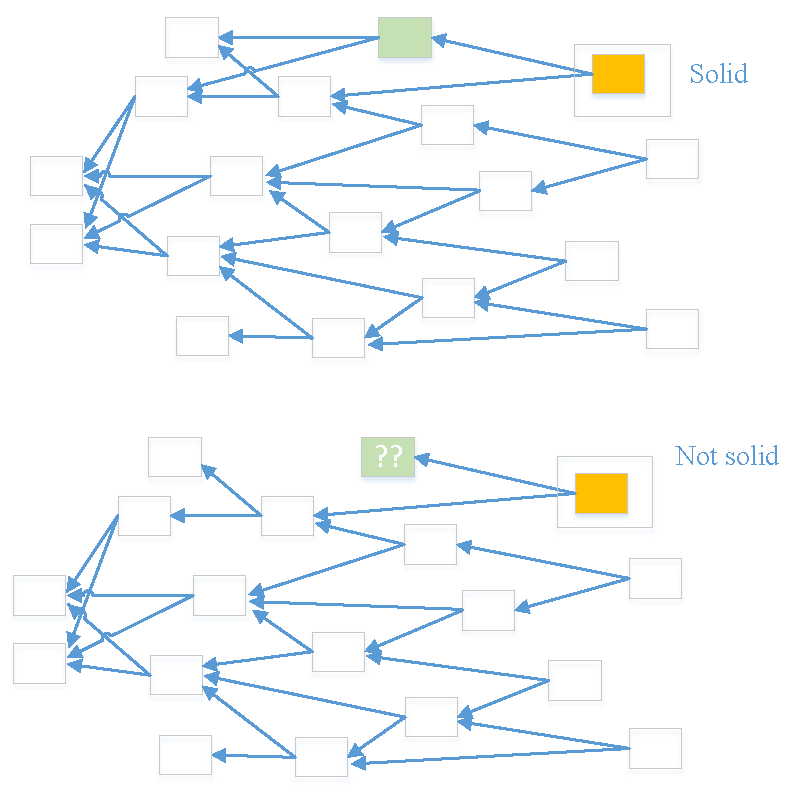}
	\caption{A comparison between solid and unsolid transactions. The yellow transaction is solid in the upper ledger, but is not solid in the lower ledger because the lower ledger does not contain the green transaction yet.}
	\label{fig: Solidification}
\end{figure}

\begin{definition}[{\em Solid transaction}]\label{def: solid}
If a node has in its ledger all transactions in the past cone of a given transaction, then we say that this transaction is solid. An example of a solid and unsolid transaction are shown in Figure \ref{fig: Solidification}.\newline
\end{definition}

\begin{definition}[{\em Disseminated transaction}]\label{def: Disseminated transaction}
A transaction is said to be disseminated when it reaches all nodes in the network.\newline
\end{definition}
\begin{definition}[{\em Dissemination rate}]\label{def: Disseminated rate}
The rate of the disseminated transactions issued by node $i$ is denoted by $DR_i$. Moreover, $DR = \sum_{i}DR_i$ denotes the total dissemination rate from all nodes.\newline
\end{definition}
\begin{definition}[{\em Latency}]\label{def: Latency}
Latency refers to the period of time between a transaction being issue and when it is added to the ledgers of all other nodes. In case a transaction is not delivered, this will have latency infinity.\newline
\end{definition}

The following definitions relate to requirements of the algorithm and are useful when we evaluate the efficacy of the access control algorithm. They are taken directly from \cite{cullenaccess}.

\begin{definition}[{\em Consistency}]\label{def: Consistency}
If a transaction issued by an honest node (i.e. a node obeying the proposed protocol) is written by one honest node, it should eventually be written by all honest nodes.
\end{definition}
\begin{definition}[{\em Fairness in dissemination rate}]\label{def: Fairness in dissemination rate}
The fairness in dissemination rate means that the dissemination rate of each node should be allocated proportional to the node's reputation. 
\end{definition}
\begin{definition}[{\em Fairness in latency}]\label{def: Fairness in latency}
The fairness in latency means that, for a given dissemination rate relative to the node’s reputation, a node’s transactions should experience similar latency. In other words, the latency of a nodes' transactions is related to reputation-scaled dissemination rate \footnote{ {\color{black} The reputation-scaled dissemination rate is the value that dissemination rate of this node divide the node's reputation value. Other terms, including reputation-scaled inbox length, are defined similarly to this.}}and not a node's own reputation.
\end{definition}
\begin{definition}[{\em Security}]\label{def: Security}
Malicious nodes that arbitrarily deviates from the proposed protocol should be unable to interfere with any of the above requirements.
\end{definition}

\section{Analysis and extensions of~\cite{cullenaccess}}
As we have already mentioned, this work builds on the access control algorithm proposed in~\cite{cullenaccess}, addressing a number of potential attack scenarios not considered in the introductory paper and considering requirements arising from the DAG structure of the ledger in more detail. We now review the existing algorithm and introduce the improvements proposed in this work.
\subsection{Access control algorithm in~\cite{cullenaccess}}

The congestion algorithm  in~\cite{cullenaccess} is organized around three functional components: a {\em scheduling} algorithm; a {\em rate setting} algorithm; and a {\em buffer management} policy.\newline 

\begin{itemize}
\item {\em Scheduling}: In order to accommodate latency-critical flows arising from bursty arrivals,~\cite{cullenaccess} proposes a modified version of the Deficit Round Robin (DRR) scheduling algorithm, DRR-- (DRR minus), where each flow is connected with a queue and is served in a round robin manner~\cite{shreedhar1996cient} according to node reputation. \newline
	
\item {\em Rate setting}: 
In order to maximize the utilization of the network and ensure that nodes are not overwhelmed, an AIMD method is adopted by each node. Specifically, every node checks the length of its inbox in order to gauge the congestion level of the network and adapts its issue rate accordingly. Relying only on local congestion measurements is enough to evaluate the degree of congestion of the network since transactions are broadcast to all network participants which schedule them at the same rate. Hence, all nodes will see the same number of transactions within a short time-frame. Unlike in TCP, where the protocol considers explicit notification transactions, the usage of local information is fundamental to properly deal with an adversarial environment.\newline

\item {\em Buffer management}: 
When the buffer becomes full, this component drops transactions depending on the number of transactions in the inbox issued by each node, weighted by the issuing node's reputation.\newline

\end{itemize}

\begin{figure}[ht]
	\centering
	\includegraphics[width=\columnwidth]{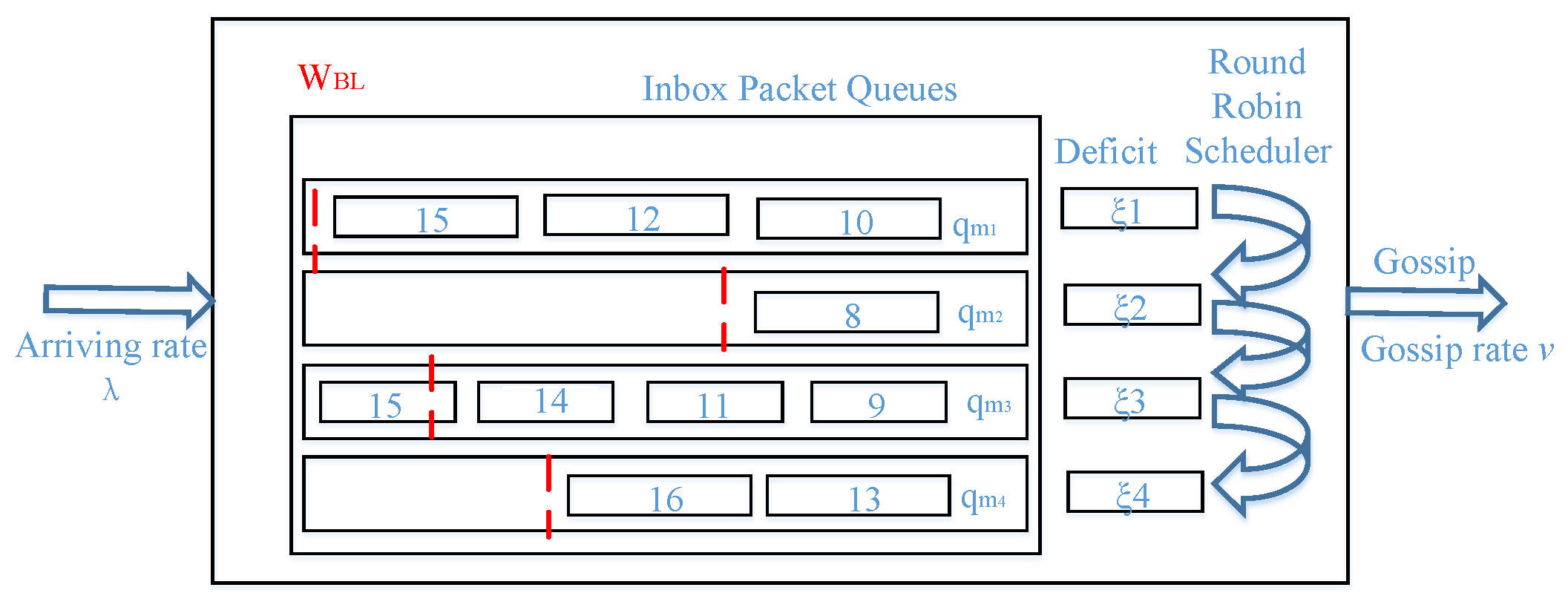}
	\caption{Inbox of node $m$. Each transaction in the queue is solid and ordered by timestamp. The dotted red line above represents the blacklisting threshold which is concerned with reputation-scaled inbox length of each node. The deficit every node gets each round is proportional to node reputation.}
	\label{fig: Networkmodel}
\end{figure}

\subsection{Extensions of access control in~\cite{cullenaccess}}

In~\cite{cullenaccess}, a baseline access control algorithm was shown to satisfy requirements and perform well in an honest environment, and demonstrated an ability to filter out the transactions of malicious nodes in the case of a simple attack scenario. However, an extensive security analysis was not performed in~\cite{cullenaccess} and the following key points were not addressed.\newline

\begin{itemize}
	\item \textit{Blacklisting}: While the buffer management scheme proposed in~\cite{cullenaccess} ensures that malicious nodes can not cause the transactions of honest nodes to be dropped, there is no explicit mechanism for \emph{blacklisting} malicious nodes to avoid wasting resources on malicious spam.\newline
	\item \textit{Solidification}: The buffer management proposed in~\cite{cullenaccess} does not take solidification of transactions within the  DAG structured ledger into consideration (see Definition \ref{def: solid}). In particular, dropping malicious transactions may prevent the solidification of subsequent honest transactions. For example, if node $i$ receives transactions from malicious nodes and other new coming transactions have already appended to the ledger, at this time dropping malicious transactions would affect the consistency of the ledger.\newline
	\item \textit{Attacks}: Only a basic attack scenario is considered in~\cite{cullenaccess}, which does not account for the ability of a malicious node to send different traffic to different neighbours, operate numerous nodes, or find new peers after being identified as an attacker by its current peers.\newline

\end{itemize}

We propose to extend the algorithm of~\cite{cullenaccess} by introducing a blacklisting mechanism and a number of modifications to explicitly take the need for solidification and the associated issues into account. These improvements ensure that more advanced attacks can be prevented, as we demonstrate by simulation in Section \ref{sec: sims}. \newline

\subsubsection{Blacklisting}

We define the reputation-scaled queue length as
\begin{equation}
    \hat{q}_{m_{i}} \triangleq \frac{q_{m_{i}} }{rep_i}, \qquad \forall i\in|\mathcal{N}|.
\end{equation}
We augment the protocol in \cite{cullenaccess} by adding a condition $\hat{q}_{m_{i}} >W_{BL}$, where $W_{BL}$ is the threshold to blacklist a node. As shown in Figure \ref{fig: Networkmodel}, when a node $m$ receives transactions from its neighbours, if the reputation-scaled queue length related to node $i$ is larger than $W_{BL}$, then node $m$ will blacklist node $i$. Consequently, it will drop any new incoming transactions issued by node $i$, and update the time when node $m$ blacklist this misbehaving node $i$ (malicious node). We denote this time by $T_{BL_{i}}$. Note that transactions that are already in the inbox of node $i$ will be scheduled as normal to avoid potential inconsistencies, as these transactions may have already been added to the local ledger by other nodes in the network.\newline

Additionally, we propose an improved AIMD rate setter. Depending on whether or not blacklisting has occurred recently, two cases are considered. If no blacklisting has happened yet, or the time since the last blacklisting even is greater than a given time threshold $\epsilon$, the rate setter follows the same AIMD algorithm as in~\cite{cullenaccess}. Namely, if the reputation-scaled queue length, $\hat{q}_{m_{m}}$ is larger than a certain threshold $W$, the issuing rate of node $m$ decreases through a multiplicative decrease parameter $\beta$\protect\footnote{Recall that the AIMD algorithm \cite{corless2016aimd} is characterised by two parameters; an additive increase parameter $\alpha > 0$, that determines the rate at which the node probes for available bandwidth; and a multiplicative decrease parameter $\beta \in (0,1)$, that determines the fraction of resource the node releases in response to congestion. In the rate setting defined in \cite{cullenaccess} these parameters depend on their reputation.}. After decreasing the rate, the process of issuing and rate setting waits for $ \tau$ seconds to allow the network to stabilise. On the other hand, if $\hat{q}_{m_m}$ is less than $W$, the issuing rate of node $m$ increases by local additive increase parameter -- the product of $\alpha_m \cdot |tx|$. $\alpha_m$ is defined as $A \cdot \frac{rep_m}{\sum{rep}}$, where $A$ is a global additive increase parameter for all nodes and $|tx|$ denotes the writing work needed for a transaction to be added to the ledger. On the other hand, if the time since the last blacklisting event is smaller than a given threshold, $\epsilon$, the issuing rate of node $m$ is set to be proportional to the issuing rate of a content node, $\tilde{\lambda}_m$, where the  parameter $\sigma$ is a given fixed value. The blacklisting algorithm is depicted in Algorithm \ref{alg: Blacklisting}.\newline

\textbf{Remark:} When a node is blacklisted, there is an instantaneous drop in local traffic, and as blacklisting can happen at slightly different times for different nodes, this can cause discrepancies between nodes' local views of traffic levels. This is particularly critical for best-effort nodes that try to fill the spare bandwidth after when traffic levels drop, which may cause them to be perceived as attackers by their neighbors that have not yet blacklisted the malicious node. For this reason, we pause the rate setting increase for some time after blacklisting.\newline

\textbf{Remark:} The blacklisting method in our work is a local method, requiring no additional global information. Local blacklisting is enough as it is likely that honest neighbours will drop an attacker's transactions, and hence, these will not be propagated through the network. In the event the attacker tries to change neighbours (i.e., re-peering) then it will be blacklisted by these new neighbours as well. In order to test the efficacy of this designed algorithm, a set of simulations are presented in the next section.\newline

\begin{algorithm}[h!]
\caption{Blacklisting algorithm }\label{alg: Blacklisting}
\begin{algorithmic}[1]
\Statex \emph{Part 1: How to blacklist a malicious node in node's inbox}
\If{Node receive a transaction issued by other nodes}
   \If {$\hat{q}_{m_{i}} > W_{BL}$}
      \State Blacklist node $i$
      \State Drop transactions issued by node $i$
      \State Update $T_{BL_{i}}$
   \EndIf
\EndIf

\Statex \emph{Part 2: Improved AIMD Rate Setter} 
\Statex \emph{Repeat each time a transaction is issued:}\If{$T_{BL_{i}}==0~or~T_{H_{i}}>\epsilon$}
    \If{$\hat{q}_{m_m}> W $} \label{line: backoff conditions}
       \State $\lambda_{m} \gets \lambda_{m} \cdot \beta$ 
        \State Stop rate setting and issuing for $ \tau$ seconds
    \Else
        \State $\lambda_{m} \gets \lambda_{m} + \alpha_{m} \cdot |tx|$
    \EndIf
\Else
        \State $\lambda_{m} \gets \sigma \cdot \tilde{\lambda}_m$
\EndIf
\end{algorithmic}
\end{algorithm}

\subsubsection{Solidification requests}

Due to the DAG structure of the ledger, and due to delays in the network, it is possible that some transactions at this point might not yet be solid (see Definition \ref{def: solid}). To avoid these kind of situations, and to maintain consistency of the ledger, we introduce a new component called {\em solidification requests} to ensure that all the transactions in the past cone are received by the node. When a transaction arrives at the front of the queue in the node's inbox, if the transaction is not solid, the node sends a solidification request to ask its neighbours to send the missing transactions in the past cone. The transaction can only be scheduled when it becomes solid (full past-cone is received). By introducing the rule that transactions should not be scheduled until they are solid, we may encounter issues when nodes are blacklisted and the transactions of the blacklisted node are required to solidify. This problem can be further exaggerated by an attacker by sending different streams of transactions to different neighbours. To deal with this problem, we also introduce ordering of transactions in the inbox based on their timestamps, this ensures that transactions are scheduled and forwarded in roughly the same order by all nodes, even if they are received in different order from malicious nodes. Solidification requests are further explored in Section~\ref{sec: sims}, where it is shown that if this component is not deployed, the dissemination rate of the whole network drops to zero. The solidification algorithm is described in Algorithm \ref{alg: Solidification}.\newline

\begin{algorithm}[h!]
\caption{Solidification Requests }\label{alg: Solidification}
\begin{algorithmic}[1]
\Statex \emph{Repeat each time a transaction $tx$ is scheduled:}
\If{$p$ is $unsolid$}
    \If{$p$ not in $ReqSolid$/$Tran$/$Scheduled$}
         \State add $p$ into $ReqSolid$
    \EndIf
\Else
    \State do normal DRR-- scheduler
\EndIf
\end{algorithmic}
\end{algorithm}




\section{Simulation results} \label{sec: sims}

\subsection{Simulation setup}
\begin{figure}[H]
\centering
\includegraphics[width=\columnwidth]{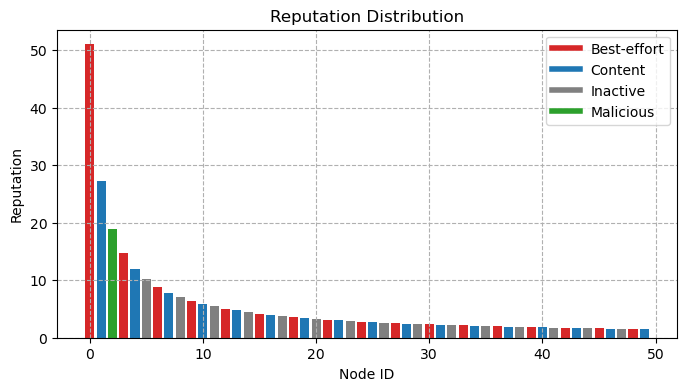}
\caption{Reputation distribution follows a Zipf distribution with exponent 0.9. As shown by each bar’s colour, nodes are Best-effort in red, Content in blue, Inactive in grey and malicious in green.}
\label{fig: RepDist1}
\end{figure}

In order to test the robustness and effectiveness of the designed algorithm under potential attack scenarios, we have built a Python simulator. In what follows, the network used to illustrate our results is composed of 50 nodes, each of which is peered with 4 randomly chosen neighbours. Communication delays between nodes are exponentially distributed in the range from 50 ms to 150 ms. The reputation distribution used in our simulator follows the measured distribution from the IOTA network and is depicted in Figure \ref{fig: RepDist1}. Specifically, we use the number of transactions issued by each account in the IOTA network, which follows a Zipf distribution\footnote{Wealth has also been shown to follow similar distributions, so this model is also well suited to reputation systems derived from wealth, i.e., PoS \cite{jones2015pareto}.} with exponent 0.9. The scheduling rate $\nu$ is set to 50 transactions per second. Other relevant parameters are set as follows: the maximum deficit for each empty queue $DC_{max} = 1$, $A = 0.06$, $\beta = 0.5$, $\tau = 2$, $\sigma = 0.6$, $|tx=1|$, $W_{BL}=5$ and $\epsilon = 15$ seconds (see Table~\ref{tab:sim-params-1}). The effect of parameters, such as increase parameter $\alpha$, the decrease parameter $\beta$, the work threshold $W$, and the total number of nodes, are shown in~\cite{cullenaccess}, the interested reader can refer to that. For each experiment, ten Monte Carlo simulations are performed.\newline

\begin{table}[ht]
	\caption{Access control algorithm parameters used in simulations.}
	\centering
	\begin{tabular}{c c || c c c c c||c||}
		\multicolumn{2}{c||}{\textbf{Scheduler}} & \multicolumn{5}{c||} {\textbf{Rate Setter}}& \multicolumn{1}{c||} {\textbf{Blacklisting}} \\
		$\nu$ & $DC_{max}$ & $A$ & $\beta$ & $\tau$ & $\sigma$ & $\epsilon$& $W_{BL}$\\
		\hline
		$50$ & $1$ & $0.06$ & $0.5$ & $2$ & $0.6$ & $15$ & $5$ \\
	\end{tabular}
	\label{tab:sim-params-1}
\end{table}

In our experiments, there are two types of malicious nodes:\newline

 \begin{itemize}
    \item The first type are malicious nodes sending above the rate allowed by the rate setter module to its neighbours. We call this a {\em spamming attacker}.\newline
    \item The second type are malicious nodes that send different streams of transactions to different neighbours, while each stream individually obeys the rate setter indications. These are named {\em multi-rate attackers}.\newline
\end{itemize}

Malicious nodes can simultaneously be both spamming and multi-rate attackers. Furthermore, we assume that attackers can change their neighbours as soon as they detect that they are blacklisted by all neighbours. We call this action ``re-peering'', and it results in a more sophisticated attack scenario.\newline

\textbf{Remark:} Multiple coordinated malicious nodes in the system at the same time constitutes a very powerful form of attack. As we shall see, our modified algorithm is able to cope with such a scenario.\newline 

The rest of the section is organized as follows. In Section~\ref{sec:attacks} we consider the following attack scenarios:\newline
\begin{itemize}
    \item The first experiment considers a spamming attacker without re-peering.\newline
	\item The second experiment considers a spamming attacker with re-peering.\newline
	\item The third experiment considers a multi-rate attacker without re-peering.\newline
    \item The fourth experiment considers multi-rate attacker with re-peering.\newline
	\item The fifth experiment considers multiple nodes attacking the network simultaneously.
\end{itemize}
Then, in Section~\ref{sec:robustness-analysis} we present an analysis of the robustness of the protocol:\newline
\begin{itemize}	
	\item The first experiment considers when the nodes' reputation varies over time.\newline
	\item The second experiment considers the impact of active nodes becoming inactive and switching back to being active.
\end{itemize}

Finally, in Section~\ref{sec:comparison} a number of experiments are presented to illustrate the improvements of our algorithm compared to~\cite{cullenaccess}. Note that to the best of our knowledge,~\cite{cullenaccess} is the first piece of work which proposes reputation-based access control for DAG-based ledgers and removes the need for PoW for Sybil protection. So here we can only provide this one comparison.
 
\subsection{Analysis of the access control algorithm in a malicious environment}\label{sec:attacks}
\subsubsection{Spamming attacker without re-peering}
\begin{figure*}
	\centering
	\subfigure[]{
		\begin{minipage}[b]{0.3\linewidth}
			\includegraphics[width=6.1cm, height=3.5cm]{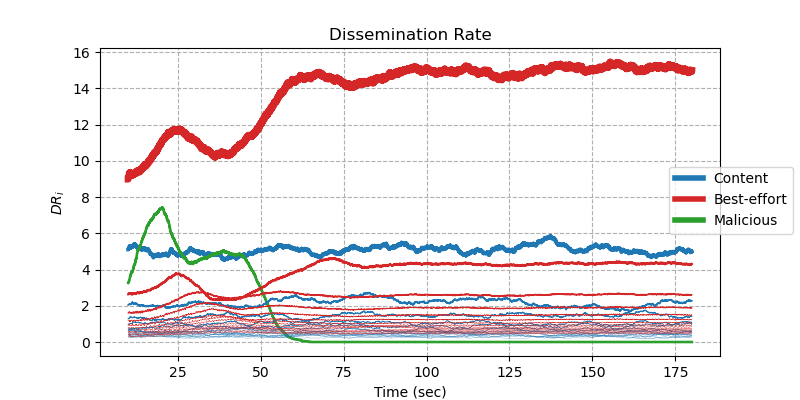} 
		\end{minipage}
		\label{fig:dissemination1}
	}
	\hspace{1mm} 
    \subfigure[]{
    	\begin{minipage}[b]{0.3\linewidth}
   		\includegraphics[width=6.1cm, height=3.5cm]{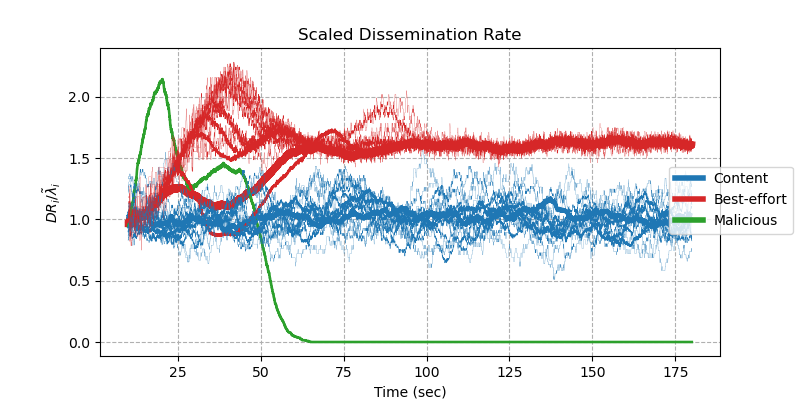} 
    	\end{minipage}
	\label{fig:Sdissemination1}
    }
    	\hspace{4mm} 
    	\subfigure[]{
		\begin{minipage}[b]{0.3\linewidth}
			\includegraphics[width=5.7cm, height=3.2cm]{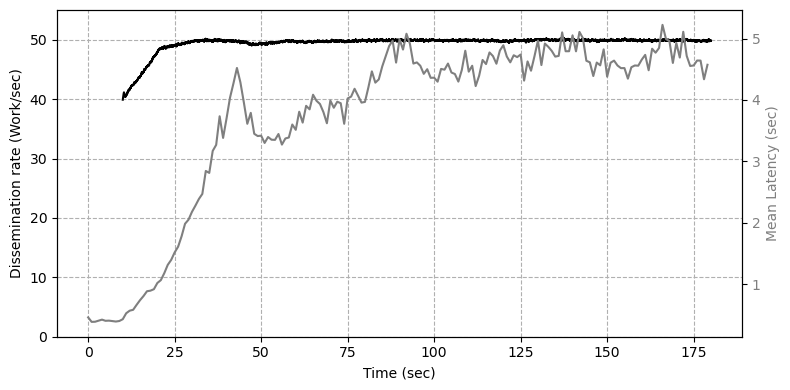}
		\end{minipage}
		\label{fig:Throughput1}

}
    \subfigure[]{
    	\begin{minipage}[b]{0.3\linewidth}
   		\includegraphics[width=5.7cm, height=3.2cm]{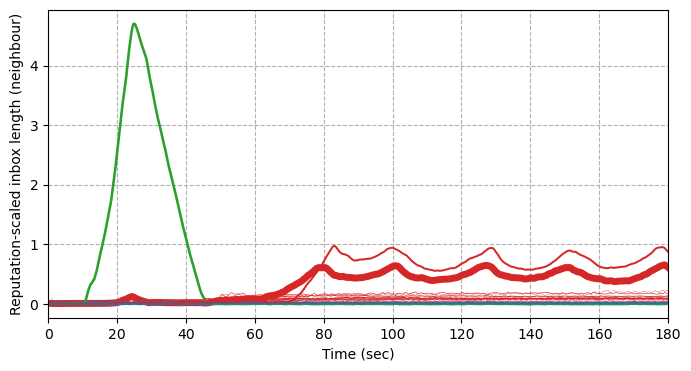} 
    	\end{minipage}
	\label{fig:AvgInboxLen1}
    }
     \subfigure[]{
    	\begin{minipage}[b]{0.3\linewidth}
   		\includegraphics[width=5.7cm, height=3.2cm]{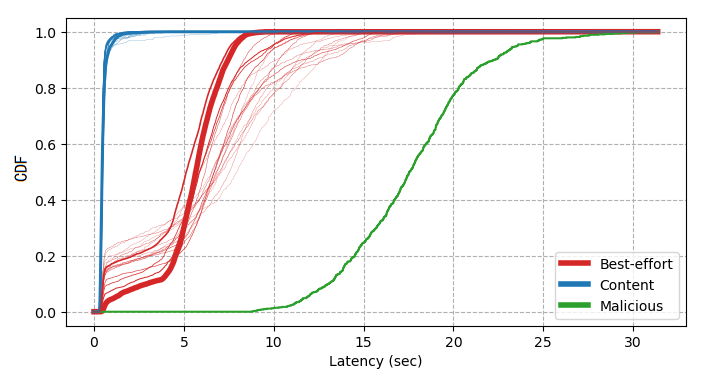} 
    	\end{minipage}
	\label{fig:Latencynew1}
    }
	\caption{This set of figures is spamming attacker scenario without re-peering. (a) is dissemination rate of each node and (b) is scaled dissemination rate of each node. (c) is dissemination rate and mean latency for each node. (d) is reputation-scaled inbox length of one randomly selected malicious node's neighbour. Transactions issued by honest nodes are in red, while transactions issued by malicious nodes are in green. (e) is the cumulative density function of latency across all transactions for all nodes.}
	\label{fig:spamming_without}
\end{figure*}

\ 
\newline
\indent In this subsection, we consider a spamming attacker which issues transactions at a larger rate that allowed by the rate setter module. This malicious node does \textit{not }reconnect with other neighbors of the network after being blacklisted. Figure \ref{fig:dissemination1} and \ref{fig:Sdissemination1} show the dissemination rate and reputation-scaled dissemination rate per node (see Definition IV.4) respectively. We use red lines to plot {\em best-effort nodes}, green lines for {\em malicious nodes} and  blue lines for {\em content nodes}. Furthermore, in the plots, the thickness of every line is chosen to be proportional to each node's reputation. The difference between Figure \ref{fig:dissemination1} and \ref{fig:Sdissemination1} is that,  \ref{fig:Sdissemination1} depicts the reputation-scaled dissemination rate. From \ref{fig:Sdissemination1}, we can see that the value of best effort nodes and content nodes converge to a constant value respectively by the end of the simulation. This means that fair access is eventually ensured for every node. Note that both the dissemination rate and the reputation-scaled dissemination rate of the attacker drop to zero since the malicious node has been blacklisted by all neighbours, hence isolated. This of course means that all transactions from this malicious node will be discarded.\newline

 
    The dissemination rate and the mean latency over all disseminated transactions are presented in Figure \ref{fig:Throughput1}. In this plot, even when attackers appear, the dissemination rate $DR$ for all transactions converges to a constant value, as does the mean latency.\newline


    The reputation-scaled inbox length of a randomly-chosen neighbor of the attacker is depicted in Figure \ref{fig:AvgInboxLen1}. As can be observed the reputation-scaled inbox length of honest nodes is low. This is especially true between 10 and 50 seconds when the spammer node is trying to fulfill the available bandwidth. This value is also low between 50 seconds to 60 seconds because we have set the malicious nodes neighbours issuing rate to be proportional to that of a content node with their $T_{{BL}_i}$ is less than 15 seconds.\newline


    The fairness in latency of the revised protocol is validated in Figure \ref{fig:Latencynew1}. In particular, Figure \ref{fig:Latencynew1} depicts the cumulative density function of the latency of the transactions issued by each node. It is clear from this plot that malicious node's transactions experience a much higher latency than other transactions: this happens because attacker's transactions cannot be scheduled by honest nodes over short time scales as they get backlogged at the inboxes of honest nodes before the attacker is blacklisted.\newline


\subsubsection{Spamming attacker with re-peering}
\begin{figure*}
	\centering
	\subfigure[]{
		\begin{minipage}[b]{0.3\linewidth}
			\includegraphics[width=6.1cm, height=3.5cm]{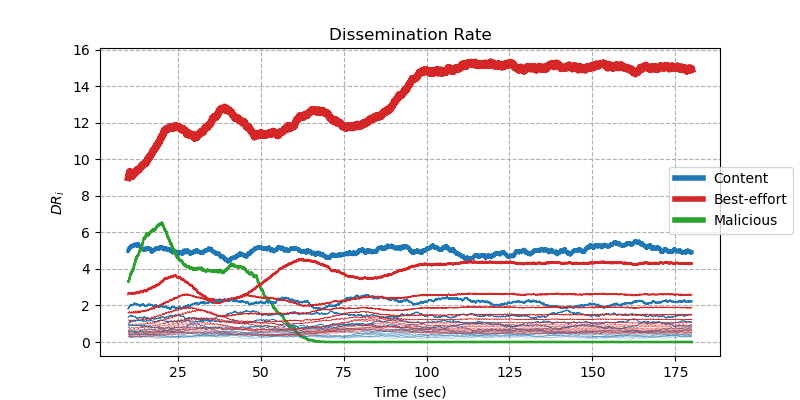} 
		\end{minipage}
		\label{fig:dissemination2}
	}
	\hspace{1mm} 
    \subfigure[]{
    	\begin{minipage}[b]{0.3\linewidth}
   		\includegraphics[width=6.1cm, height=3.5cm]{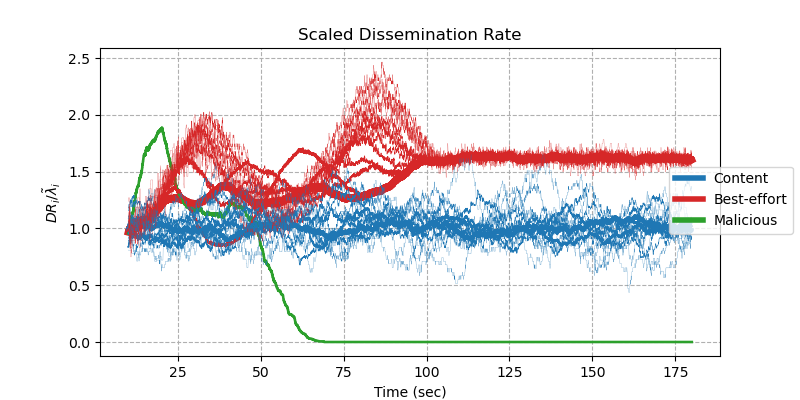} 
    	\end{minipage}
	\label{fig:Sdissemination2}
    }
    \hspace{4mm} 
    	\subfigure[]{
		\begin{minipage}[b]{0.3\linewidth}
			\includegraphics[width=5.7cm, height=3.2cm]{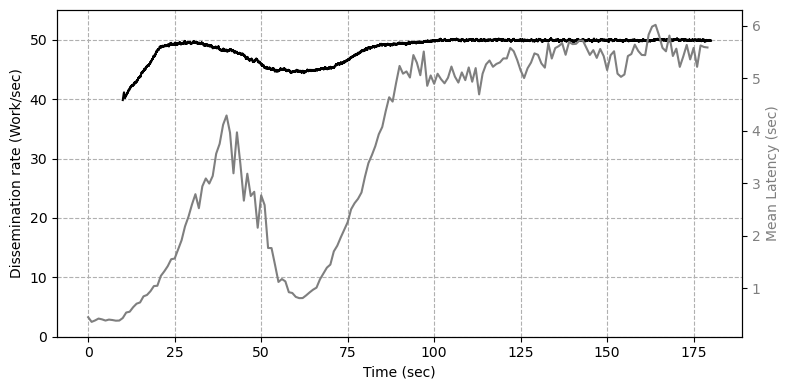}
		\end{minipage}
		\label{fig:Throughput2}

}
    \subfigure[]{
    	\begin{minipage}[b]{0.3\linewidth}
   		\includegraphics[width=5.7cm, height=3.2cm]{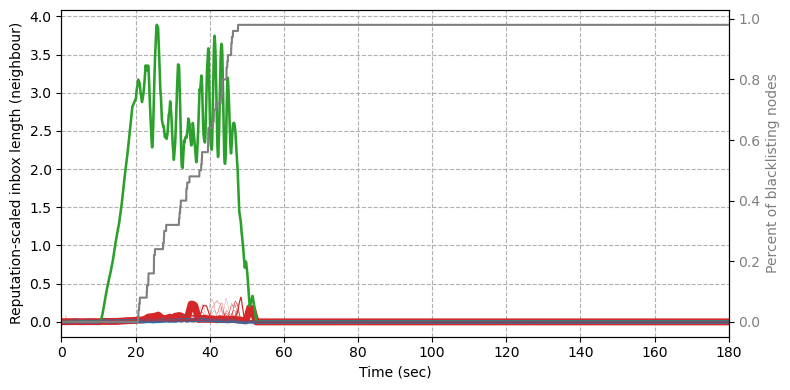} 
    	\end{minipage}
	\label{fig:AvgInboxLen1_blacklisted}
    }
     \subfigure[]{
    	\begin{minipage}[b]{0.3\linewidth}
   		\includegraphics[width=5.7cm, height=3.2cm]{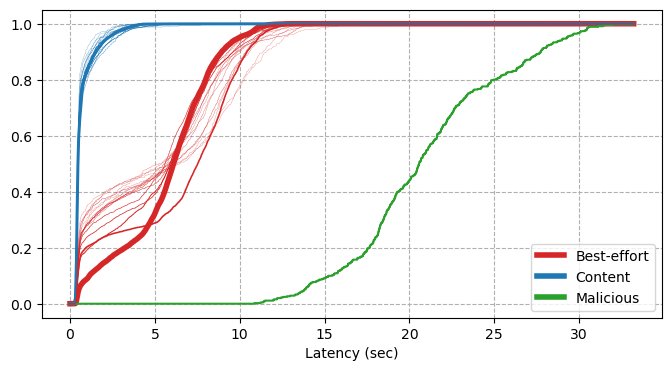} 
    	\end{minipage}
	\label{fig:Latency2}
    }
	\caption{This set of figures is spamming attacker scenario with re-peering. (a) is dissemination rate of each node and (b) is scaled dissemination rate of each node. (c) is dissemination rate and mean latency for each node. (d) is reputation-scaled inbox length of one randomly selected malicious node's neighbour. Transactions issued by honest nodes are in red, while transactions issued by malicious nodes are in green. The grey line depicts the percentage of how many how many honest nodes have been blacklisted malicious node along with time. (e) is the cumulative density function of latency across all transactions for all nodes.}
	\label{fig:spamming_with}
\end{figure*}

\ 
\newline
\indent We now consider the impact of spamming attacker with re-peering after blacklisting. As can be observed, although there are some fluctuations in the scaled dissemination rate caused by malicious nodes re-peering, the dissemination rate converges quickly to a constant value (see Figure \ref{fig:dissemination2} and \ref{fig:Sdissemination2}). Note also that the dissemination rate of malicious nodes declines quickly to zero. Thus it can be observed that the designed protocol in this paper is also effective in mitigating this type of attack.\newline 


Figure \ref{fig:Throughput2} depicts the total dissemination rate and the mean latency across all transactions for all nodes, including honest nodes and malicious nodes. Observe that there is a clear decreasing trend for both dissemination rate and latency from time 25 seconds to time 60 seconds. This is due to the gradual blacklisting of the malicious node by all honest nodes (remember that re-peering is enabled). Furthermore, it is interesting to see that malicious node is isolated, the unused bandwidth which was ``wasted'' by the attacker becomes now available and best-effort nodes can use it.\newline


The reputation-scaled inbox length of a randomly-chosen neighbor of the attacker is depicted in Figure \ref{fig:AvgInboxLen1_blacklisted}. As can be seen, honest nodes blacklist malicious node gradually because of malicious node re-peering behaviour. The grey line shows the percentage of honest nodes that have blacklisted the malicious node over time. The reason why the percentage is not at one hundred is that the malicious node can not blacklist itself. \newline

 
    The fairness in latency of the revised protocol is validated in Figure \ref{fig:Latency2}. In particular, Figure \ref{fig:Latency2} depicts the cumulative density function of the latency of the transactions issued by each node. It is same as the previous scenario that malicious node's transactions experience a much higher latency than other transactions.\newline
     
\subsubsection{Multi-rate attacker without re-peering}
\begin{figure*}
	\centering
	\subfigure[]{
		\begin{minipage}[b]{0.3\linewidth}
			\includegraphics[width=6.1cm, height=3.5cm]{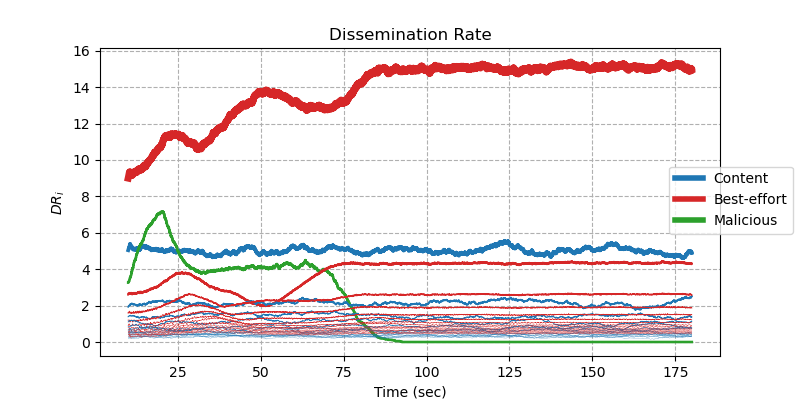} 
		\end{minipage}
		\label{fig:dissemination3}
	}
	\hspace{1mm} 
    \subfigure[]{
    	\begin{minipage}[b]{0.3\linewidth}
   		\includegraphics[width=6.1cm, height=3.5cm]{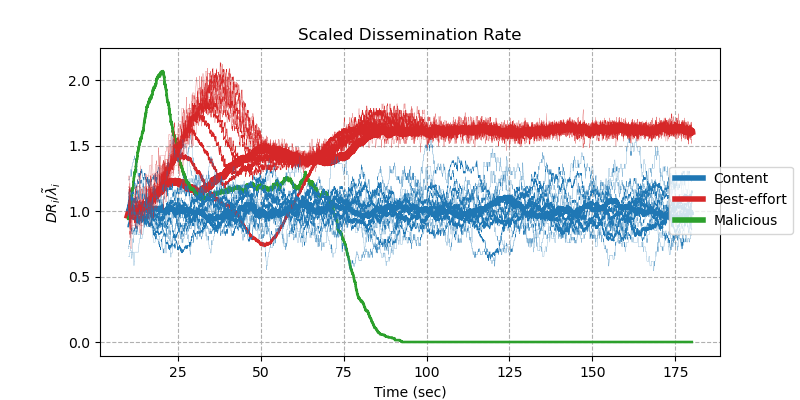} 
    	\end{minipage}
	\label{fig:Sdissemination3}
    }
    \hspace{4mm} 
     \subfigure[]{
    	\begin{minipage}[b]{0.3\linewidth}
   		\includegraphics[width=5.7cm, height=3.2cm]{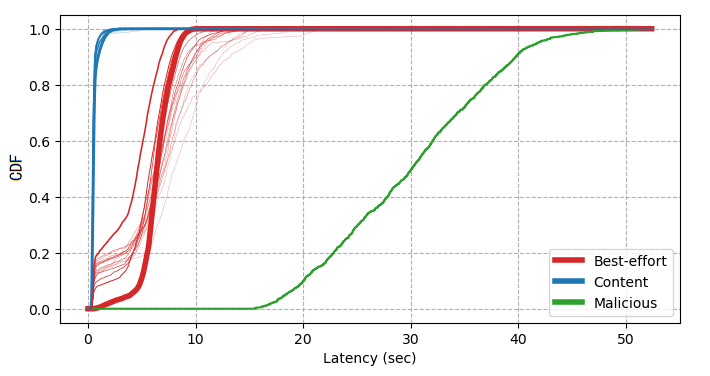} 
    	\end{minipage}
	\label{fig:Latencynew3}
    }
	\caption{This set of figures is multi-rate attacker scenario without re-peering. (a) is dissemination rate of each node and (b) is scaled dissemination rate of each node. (c) is the cumulative density function of latency across all transactions for all nodes.}
	\label{fig:Multi-rate_without}
\end{figure*}

\ 
\newline
\indent We now consider the impact of a multi-rate attacker without re-peering. The dissemination rate and reputation-scaled dissemination rate are shown in Figure \ref{fig:dissemination3} and \ref{fig:Sdissemination3}. In Figure \ref{fig:Sdissemination3}, it can observed that the fairness of dissemination rate is maintained for honest nodes, based on their reputation, throughout the simulation. While malicious nodes steal allocation from other nodes at the beginning of simulation,  this effect reduces to zero once malicious nodes spam their neighbours and they are blacklisted.\newline


Figure \ref{fig:Latencynew3} depicts the latency fairness properties of the network in this scenario. Note that when compared to content and best effort node, malicious nodes experience a much higher latency. The reason for this is as discussed above.\newline


\subsubsection{Multi-rate attacker with re-peering}
\begin{figure*}
	\centering
    	\subfigure[]{
		\begin{minipage}[b]{0.3\linewidth}
			\includegraphics[width=5.7cm, height=3.2cm]{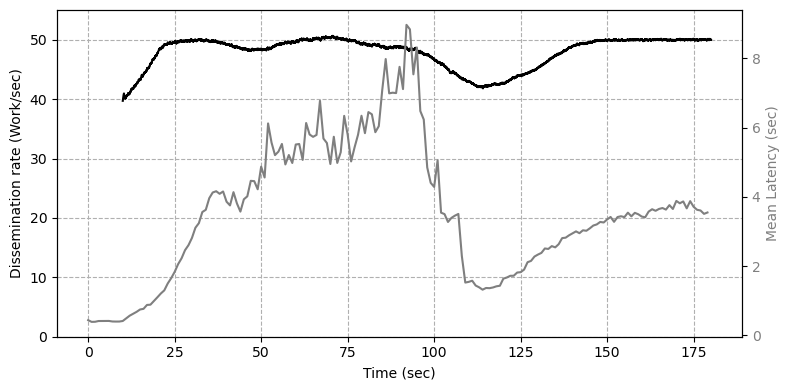}
		\end{minipage}
		\label{fig:ThroughputMRatR}

}
    \subfigure[]{
    	\begin{minipage}[b]{0.3\linewidth}
   		\includegraphics[width=5.7cm, height=3.2cm]{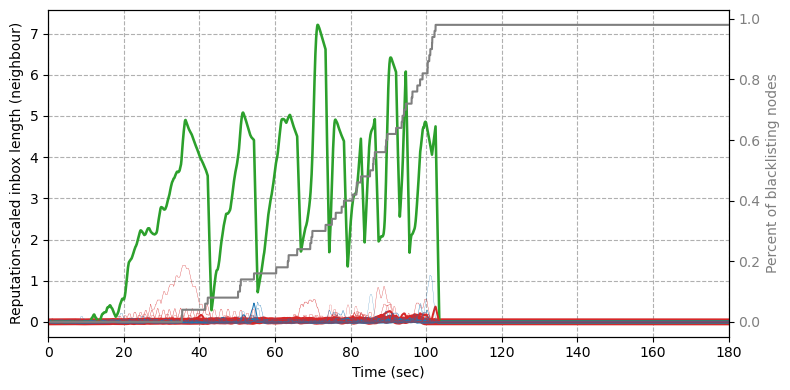} 
    	\end{minipage}
	\label{fig:AvgInboxLen2_blacklisted}
    }
    \subfigure[]{
    	\begin{minipage}[b]{0.3\linewidth}
   		\includegraphics[width=5.7cm, height=3.2cm]{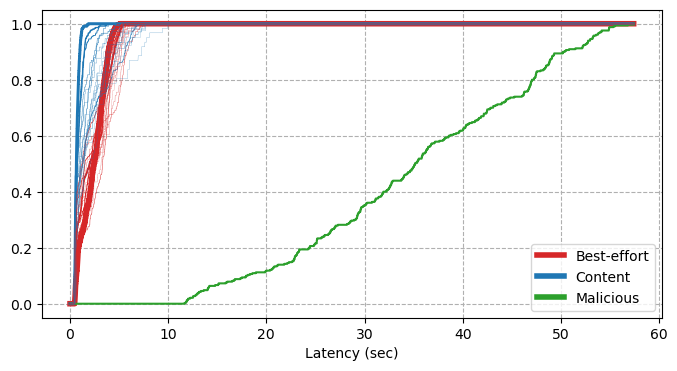} 
    	\end{minipage}
	\label{fig:Latency4}
    }
	\caption{This set of figures is multi-rate attacker scenario with re-peering.  (a) is dissemination rate and mean latency for each node. (b) is reputation-scaled inbox length of one randomly selected malicious node's neighbour. Transactions issued by honest nodes are in red, while transactions issued by malicious nodes are in green. The grey line depicts the percentage of how many how many honest nodes have been blacklisted malicious node along with time. (c) is the cumulative density function of latency across all transactions for all nodes.}
	\label{fig:spamming_with}
\end{figure*}

\ 
\newline
\indent We now consider the impact of a multi-rate attacker with re-peering. 
%
Figure \ref{fig:ThroughputMRatR} depicts the total dissemination rate and the mean latency across all transactions. Similarly to the previous experiments, attacker temporarily uses resources which it should not be using. Once the attacker is blacklisted, best-effort nodes rapidly detect the unused bandwidth and increase their throughput accordingly (from time 100 seconds onwards).\newline


The reputation-scaled inbox length of a randomly-chosen neighbor of the attacker is depicted in Figure \ref{fig:AvgInboxLen2_blacklisted}. As can be seen, due to the complexity of multi-rate attacks, especially with re-peering, the blacklisting process takes more time than in previous scenarios. The grey line shows the percentage of honest nodes that have blacklisted the malicious node over time. The reason why the percentage does not achieve 100\% is same as the spamming attack with re-peering. \newline

 
Figure \ref{fig:Latency4} depicts the latency fairness properties of the network in this scenario. Note that when compared to content and best effort node, malicious nodes experience a much higher latency. The reason for this is as discussed above. Because of the complexity of the multi-rate attack, the latency of malicious node is much higher than other scenarios.\newline
 
\subsubsection{Multiple attackers} 
\begin{figure*}
	\centering
	\subfigure[]{
		\begin{minipage}[b]{0.3\linewidth}
			\includegraphics[width=5.7cm, height=3.2cm]{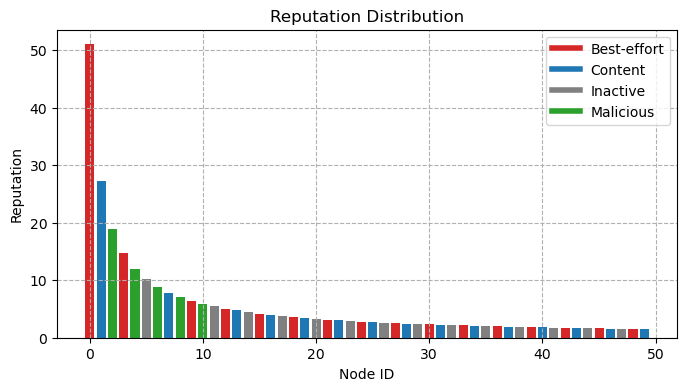} 
		\end{minipage}
		\label{fig:RepDist41}
	}	
	\hspace{-1mm} 
	\subfigure[]{
		\begin{minipage}[b]{0.3\linewidth}
			\includegraphics[width=6.1cm, height=3.5cm]{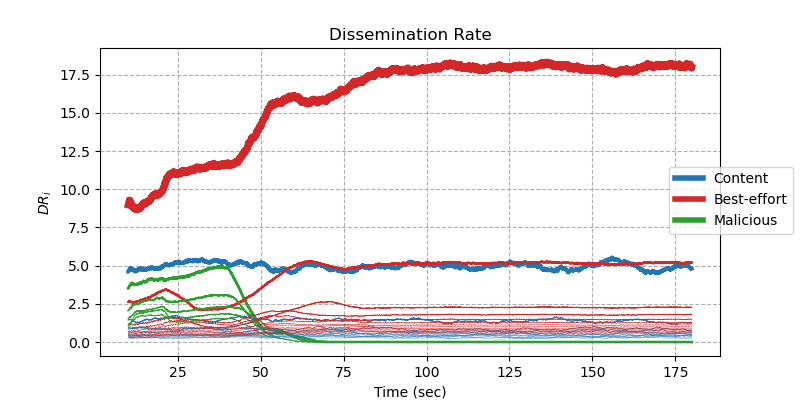} 
		\end{minipage}
		\label{fig:dissemination51}
	}
	\hspace{1mm} 
    \subfigure[]{
    	\begin{minipage}[b]{0.3\linewidth}
   		\includegraphics[width=6.1cm, height=3.5cm]{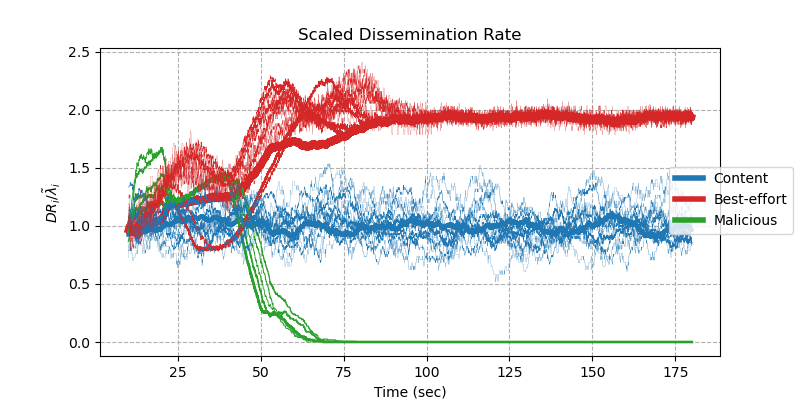} 
    	\end{minipage}
	\label{fig:Sdissemination51}
    }
    \subfigure[]{
    	\begin{minipage}[b]{0.3\linewidth}
   		\includegraphics[width=5.7cm, height=3.2cm]{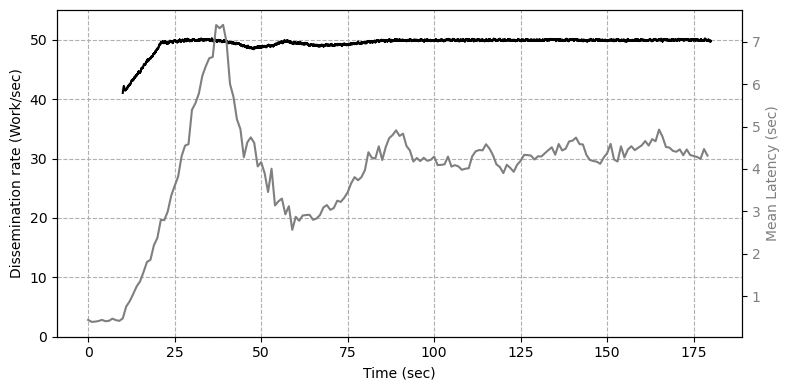} 
    	\end{minipage}
	\label{fig:Throughput51}
    }
    \subfigure[]{
    	\begin{minipage}[b]{0.3\linewidth}
   		\includegraphics[width=5.7cm, height=3.2cm]{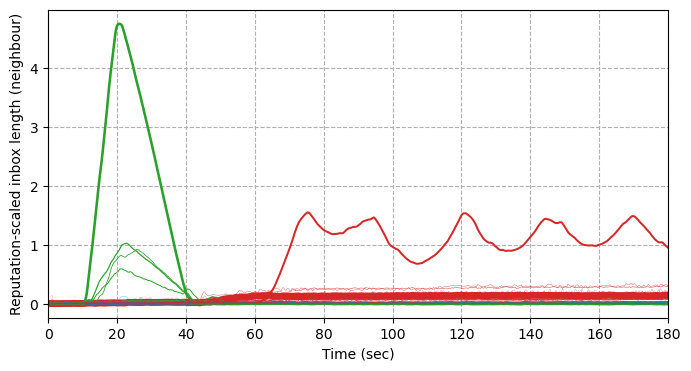} 
    	\end{minipage}
	\label{fig:AvgInboxLen41}
    }
	\caption{This set of figures is multiple spamming attackers scenario. (a) is reputation distribution follows a Zipf distribution with exponent 0.9. As shown by each bar’s colour, nodes are Best-effort in red, Content in blue, Inactive in grey and malicious in green. (b) is dissemination rate of each node and (c) is scaled dissemination rate of each node. (d) is dissemination rate and mean latency for each node. (e) is reputation-scaled inbox length of one randomly selected malicious node's neighbour.}
	\label{fig:Multiple_spamming}
\end{figure*}

\ 
\newline
\indent We now discuss the scenario when multiple malicious nodes simultaneously attack the network. First, we consider an attack performed by five spamming attackers (without re-peering). The reputation distribution of nodes of this experiment is illustrated in Figure \ref{fig:RepDist41}. As the reader can see, we choose the malicious nodes to be in the top ten nodes by reputation, which means that the a large portion of the total reputation is controlled by malicious entities, making this an exceptionally powerful attack.\newline


The fairness of dissemination rate and scaled dissemination rate is depicted in Figure \ref{fig:dissemination51} and \ref{fig:Sdissemination51}. Because we have several malicious nodes who are blacklisted by their neighbours at different times, there are fluctuations in the scaled dissemination rate of honest nodes. As it can be observed, the dissemination rate and scaled dissemination rate of malicious nodes approaches zero after they have been blacklisted.\newline


 Figure \ref{fig:Throughput51} depicts the total dissemination rate and the mean latency across all transactions. Although there is a slight fluctuation when multiple spamming malicious nodes are blacklisted by their neighbours, honest nodes start to issue more transactions to occupy the rest of the bandwidth.\newline

The reputation-scaled inbox length at a randomly-chosen neighbour of the largest-reputation attacker is depicted in Figure \ref{fig:AvgInboxLen41}. The reputation-scaled inbox length of this attacker rapidly increases up to the blacklisting threshold, after which the neighbour starts dropping attacker's transactions. Conversely, honest nodes' inbox occupations remain low up to time 60 seconds; after that, when all attackers are blacklisted, best-effort nodes can exploit the bandwidth remained unused and reputation-scaled inbox length slightly increases.\newline


\begin{figure*}
	\centering
	\subfigure[]{
		\begin{minipage}[b]{0.3\linewidth}
			\includegraphics[width=5.7cm, height=3.2cm]{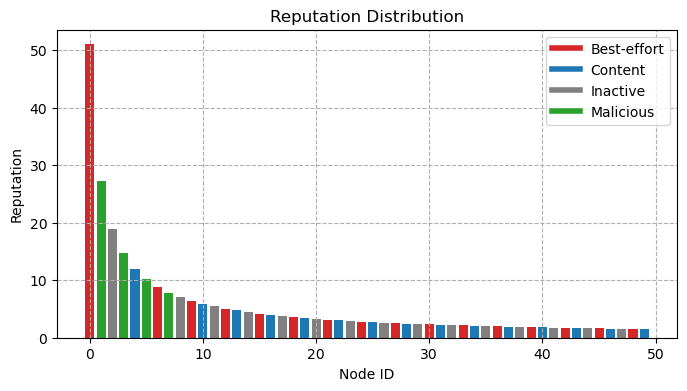} 
		\end{minipage}
		\label{fig:RepDist42}
	}
	\hspace{-1mm} 
	\subfigure[]{
		\begin{minipage}[b]{0.3\linewidth}
			\includegraphics[width=6.1cm, height=3.5cm]{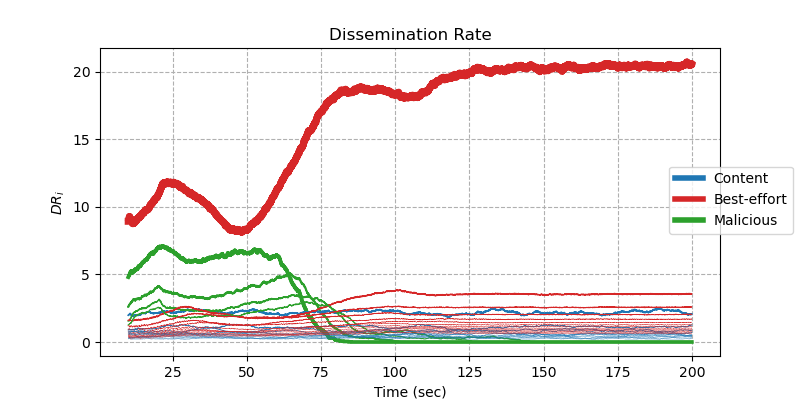} 
		\end{minipage}
		\label{fig:Dissmenation52}
	}
	 \hspace{1mm} 
    \subfigure[]{
    	\begin{minipage}[b]{0.3\linewidth}
   		\includegraphics[width=6.1cm, height=3.5cm]{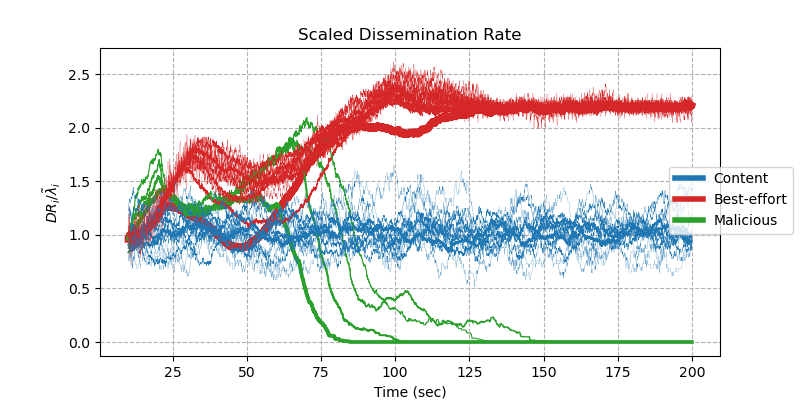} 
    	\end{minipage}
	\label{fig:SDissmenation52}
    }
    \subfigure[]{
    	\begin{minipage}[b]{0.3\linewidth}
   		\includegraphics[width=5.7cm, height=3.2cm]{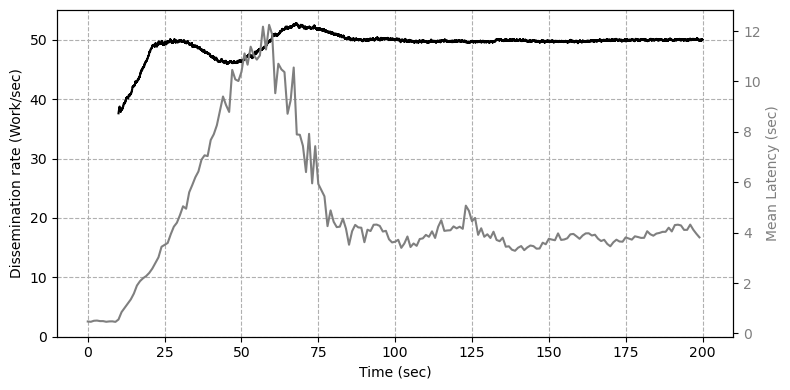} 
    	\end{minipage}
	\label{fig:Throughput52}
    }
    \subfigure[]{
    	\begin{minipage}[b]{0.3\linewidth}
   		\includegraphics[width=5.7cm, height=3.2cm]{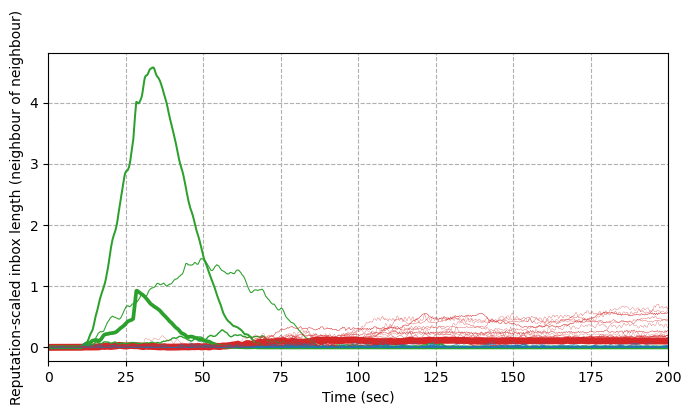} 
    	\end{minipage}
	\label{fig:AvgInboxLenMA52}
    }
	\caption{This set of figures is multiple multi-rate attackers scenario. (a) is reputation distribution follows a Zipf distribution with exponent 0.9. As shown by each bar’s colour, nodes are Best-effort in red, Content in blue, Inactive in grey and malicious in green. (b) is dissemination rate of each node and (c) is scaled dissemination rate of each node. (d) is dissemination rate and mean latency for each node. (e) is reputation-scaled inbox length of one randomly selected malicious node's neighbour of neighbour.}
	\label{fig:Multiple_spamming}
\end{figure*}

Second, we consider five multi-rate attackers trying to harm the network simultaneously. The reputation distribution in this case is illustrated in Figure \ref{fig:RepDist42}. As in the previous case, attackers are chosen among the top ten nodes by reputation. Multi-rate attackers provide a more sophisticated way to harm the network and, at the same time, more difficult to detect. The goal of a DLT is to issue a distributed database where all nodes agree on which transactions are in the ledger. When attackers send different streams of transactions to different nodes, they are trying to violate the fairness criterion and using a larger share of transactions than the one that should be guaranteed by their reputation. In our mechanism, a fundamental tool to detect this attack is provided by the fact that the scheduler sorts transactions in the inbox by timestamps: this provides an objective rule that is useful in times of congestion to let all nodes schedule (approximately) the same transactions.\newline

Figure~\ref{fig:Dissmenation52} and \ref{fig:SDissmenation52} show the dissemination rate and the reputation-scaled dissemination rate under this attack. We can verify that, even in this case, the attack is successfully repelled. The reputation-scaled inbox length at the a randomly-chosen neighbour of neighbour of the attacker is depicted in Figure \ref{fig:AvgInboxLenMA52}.\newline

Figure \ref{fig:Throughput52} depicts the total dissemination rate and the mean latency across all transactions. As in the multiple spamming attack scenario, although there is a slight fluctuation when multiple spamming malicious nodes are blacklisted by their neighbours, honest nodes start to issue more transactions to occupy the rest of the bandwidth.\newline
     
The reputation-scaled inbox length at a randomly-chosen neighbour of the largest-reputation attacker is depicted in Figure \ref{fig:AvgInboxLenMA52}. It is clear that, compared to multiple spamming attacks, multiple multi-rate attacks experience more time and each attacker may increases up to the blacklisting threshold at different time. The similar point is that, after that, when all attackers are blacklisted, best-effort nodes can exploit the bandwidth remained unused and reputation-scaled inbox length slightly increases.\newline

%

\subsection{Robustness analysis}\label{sec:robustness-analysis}
We now present a brief discussion to highlight the robustness of our protocol against a set of realistic scenarios. While the experiments in Section~\ref{sec:attacks} concern a static scenario, in this section we consider that nodes' reputation will change over time, new nodes will join the network and some of the existing nodes will change their status (from inactive to best-effort to content, and so on).\newline

\subsubsection{Time varying reputation}
\begin{figure*}
	\centering
    \subfigure[]{
    	\begin{minipage}[b]{0.3\linewidth}
   		\includegraphics[width=6.1cm, height=5cm]{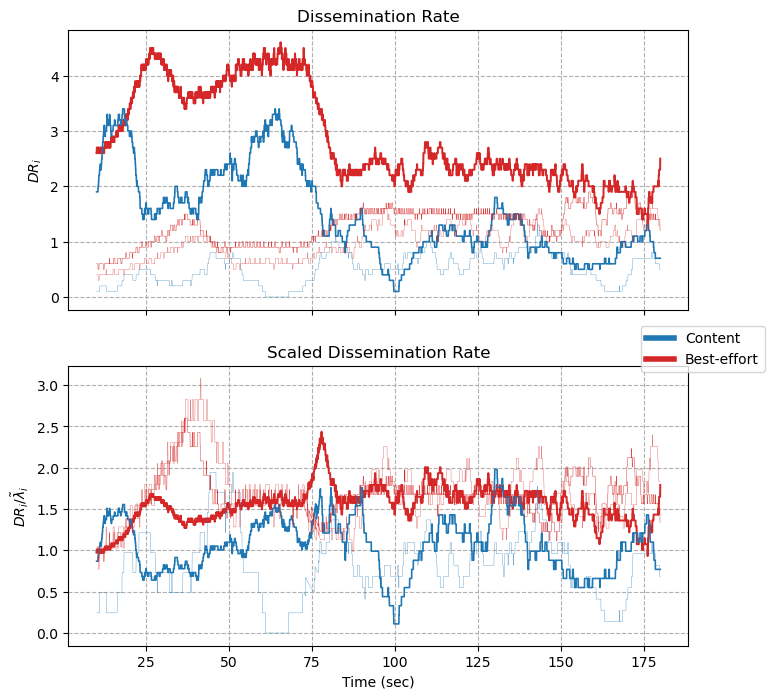}
    	\end{minipage}
	\label{fig:Part_Rates6}
    }
    	\subfigure[]{
		\begin{minipage}[b]{0.3\linewidth}
			\includegraphics[width=6.1cm, height=5cm]{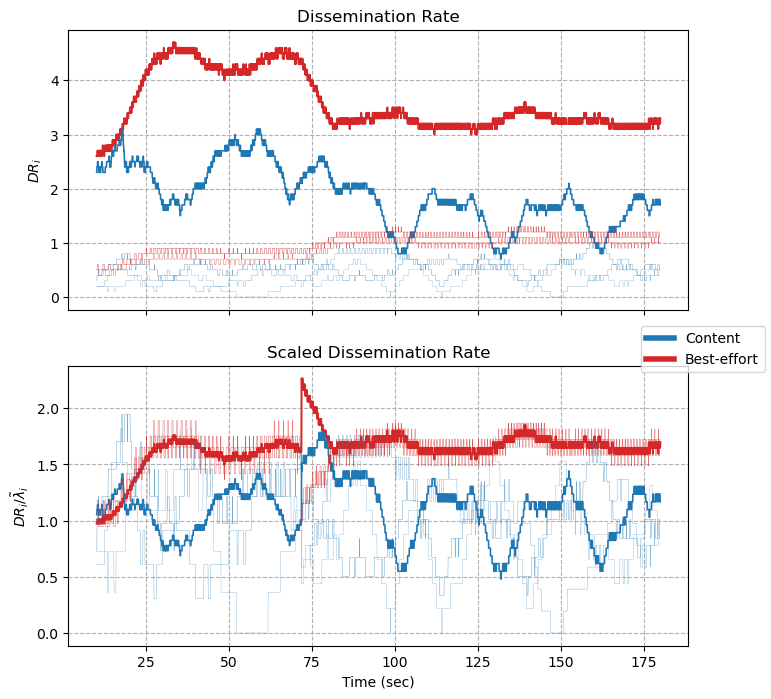}
		\end{minipage}
		\label{fig:Part_Rates_di4}
	}
		\subfigure[]{
		\begin{minipage}[b]{0.3\linewidth}
			\includegraphics[width=6.1cm, height=5cm]{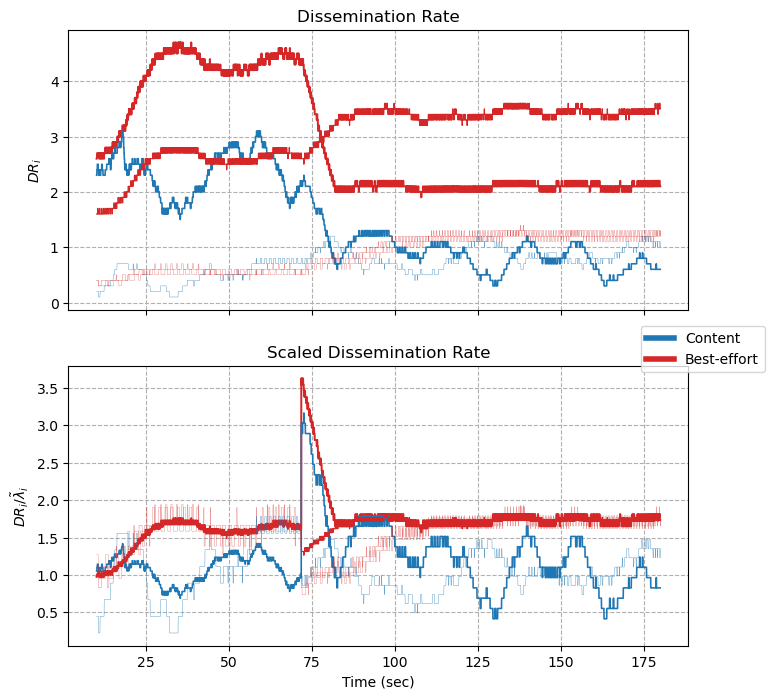}
		\end{minipage}
		\label{fig:Part_Rates_di8}
	}
    	\subfigure[]{
		\begin{minipage}[b]{0.3\linewidth}
			\includegraphics[width=6.3cm, height=3.3cm]{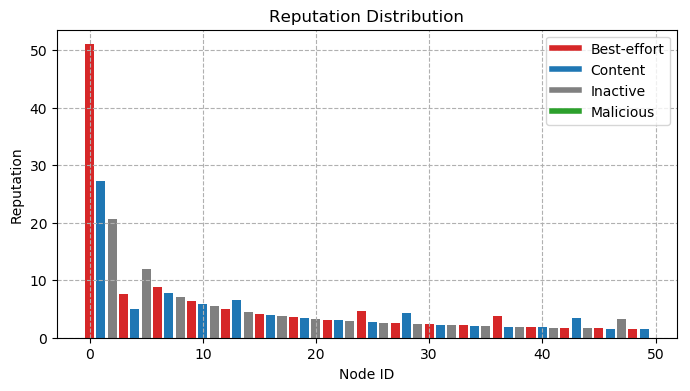}
		\end{minipage}
		\label{fig:RepDist6}
	}
		\hspace{7mm} 
    \subfigure[]{
    	\begin{minipage}[b]{0.3\linewidth}
   		\includegraphics[width=6.3cm, height=3.3cm]{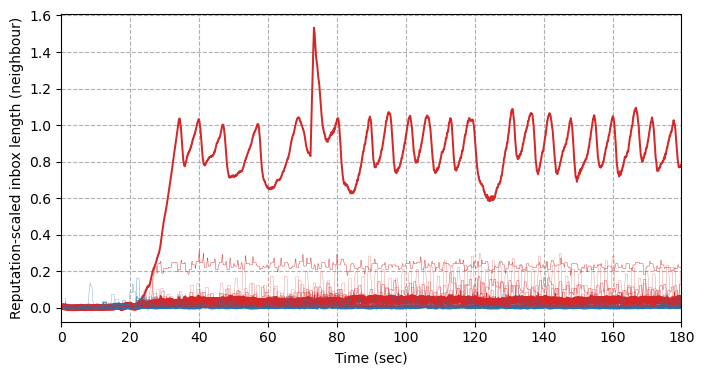}
    	\end{minipage}
	\label{fig:AvgInboxLen_rep}
    }
	\caption{This set of figures is time varying reputation scenario. (a), (b) and (c) are dissemination rate and scaled dissemination rate of each node. (d) is reputation distribution follows a Zipf distribution with exponent 0.9. As shown by each bar’s colour, nodes are Best-effort in red, Content in blue, Inactive in grey and malicious in green. (e) is reputation-scaled inbox length of one randomly selected node two’s neighbour.}
	\label{fig:Reputation_change}
\end{figure*}

We consider the impact of a change in reputation of randomly selected nodes. We randomly select two nodes and change their reputation as follows: after $75$ seconds, we decrease the reputations of nodes $3$ and $4$ (resp. best-effort and content) by $70\%$, and we increase (slightly) the reputation of their neighbors. The changes in reputation are depicted in Figure \ref{fig:RepDist6} and with specific reference to Figure \ref{fig:RepDist42}. The difference in these two figures depicts the change of each node's reputation in this experiment. Note that in this experiment we are only interested in verifying how the variation in reputation affects the operation of protocol. Consequently, it is not important which nodes are selected. The experiment is performed in an honest environment.\newline


        For this specific change in reputation, the dissemination rate, and the scaled dissemination rate of nodes 3 and 4, as well as their neighbours is depicted in Figure \ref{fig:Part_Rates6}. It can be clearly observed that there is a decreasing trend in rate for the best-effort node 3 and the content node 4 from about 75 seconds onwards. The increasing dissemination rate of the neighbours of nodes 3 and 4 is marginal. This is because each of them only acquire a small fraction of newly available bandwidth. Note also that although there is a clear spike in dissemination rate when the node reputation changes, the network quickly settles to a new steady state.\newline

The reputation-scaled inbox length of the one randomly-chosen neighbor of the attacker is depicted in Figure \ref{fig:AvgInboxLen_rep}. As we shall see, although the reputation of several honest nodes are changed during the process, there is not any risk to the system. To be specific, no honest nodes will be blacklisted due sudden loss or gain of reputation.\newline

Two more sets of simulations are presented here to illustrate the robustness and effectiveness of the designed protocol. The percentage decrease in reputation for the two nodes as $40\%$ and the dissemination rate and scaled dissemination rate of each node are depicted in Figure \ref{fig:Part_Rates_di4}. For a decrease of $80\%$ in reputation the dissemination rate and scaled dissemination rate of each node are depicted in Figure \ref{fig:Part_Rates_di8}.\newline



\subsubsection{Active nodes becoming inactive and reverting active}
We now consider the impact of nodes going offline and coming back to the network. At time 100 seconds, several best effort nodes, and several content nodes, leave the network, which means they stop issuing transactions. Then at time 200 seconds, these nodes join the network again and restart generating transactions. The dissemination rate and scaled dissemination rate of each node is depicted in Figure \ref{fig:ChangeRate1} and \ref{fig:ChangeSRate1}. Although the transient fairness of the dissemination rate of some nodes is slightly affected during the interval 100-200 seconds, the value of  scaled dissemination rate of each node converges after 200 seconds and achieves fairness eventually. 

\subsection{Comparison with protocol in~\cite{cullenaccess}}\label{sec:comparison}
Finally, to conclude the paper, in this experiment, we compare the revised protocol proposed in this paper with the unmodified one that is described in~\cite{cullenaccess}.  In~\cite{cullenaccess}, all transactions are ordered by the order they arrive at node's inbox and a simple DRR-- scheduling algorithm is employed to scheduling these transactions in node's inbox without considering the completeness of the ledger in each node. The problem with this mechanism is that the consistency can not be ensured when attacks occur.\newline

\subsubsection{Basic protocol in~\cite{cullenaccess} without solidification component} 

 The network topology and reputation distribution are as described in scenario A.(1) above. Note, in this experiment, the access control protocol contains no solidification component. The dissemination rate and scaled dissemination rate are depicted in Figure \ref{fig:ContraRate1} and \ref{fig:ContraSRates2}. As can be observed, the dissemination rate and scaled dissemination rate of many honest nodes goes to zero over time, which means that no transaction is being scheduled by all nodes. This is clearly not desirable. The reason behind this outcome is that, when there are no solidification requests, the ledger of each node may contain different transactions. This means different transactions may reach different nodes, but a transaction is only considered to be disseminated if it reaches \emph{all} honest nodes. So the consistency is not achieved when attacker sends different transactions to different neighbours under the algorithm in~\cite{cullenaccess}.\newline 

\subsubsection{The protocol in~\cite{cullenaccess} without timestamp ordering component} 
\begin{figure*}
	\centering
	\subfigure[]{
		\begin{minipage}[h]{0.3\linewidth}
			\includegraphics[width=6.1cm, height=3.5cm]{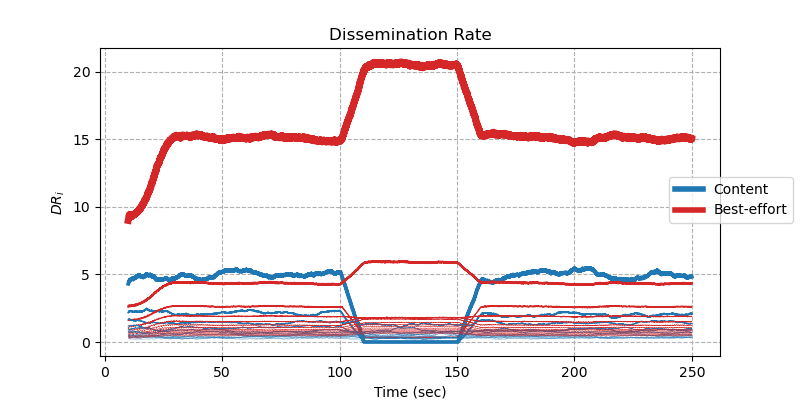}
		\end{minipage}
		\label{fig:ChangeRate1}
	}
    \subfigure[]{
    	\begin{minipage}[h]{0.3\linewidth}
   		\includegraphics[width=6.1cm, height=3.5cm]{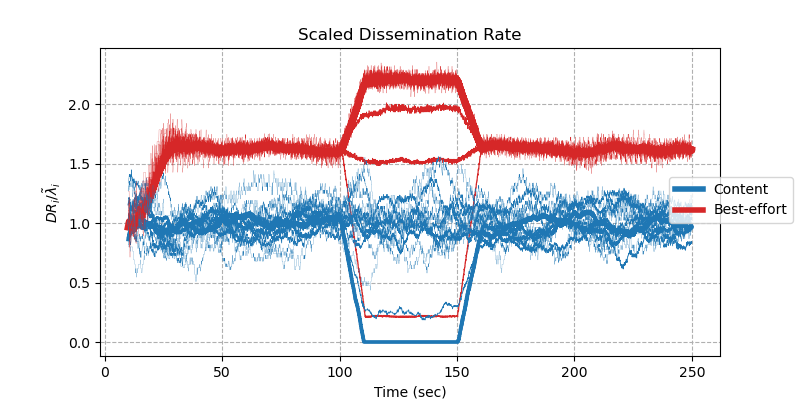}
    	\end{minipage}
	\label{fig:ChangeSRate1}
    }	
	\subfigure[]{
		\begin{minipage}[h]{0.3\linewidth}
			\includegraphics[width=6.1cm, height=3.5cm]{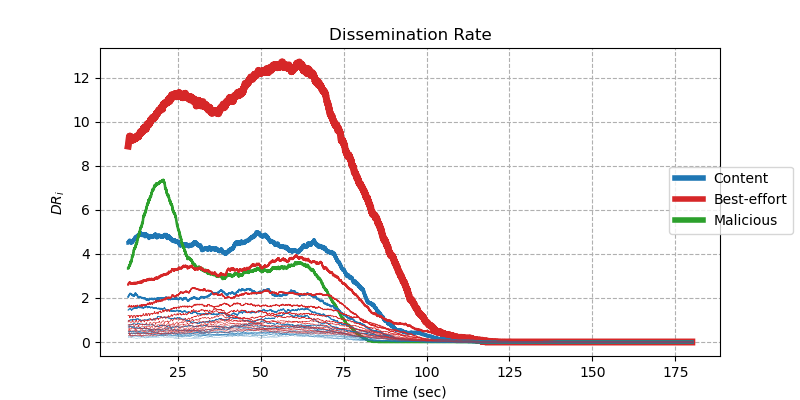}
		\end{minipage}
		\label{fig:ContraRate1}
	}
    \subfigure[]{
    	\begin{minipage}[h]{0.3\linewidth}
   		\includegraphics[width=6.1cm, height=3.5cm]{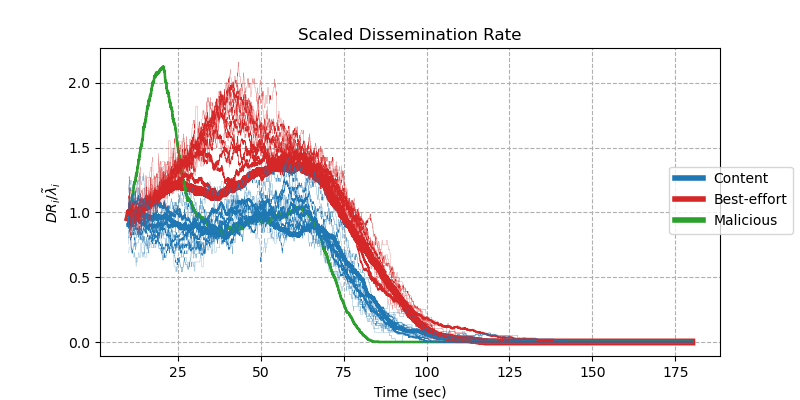}
    	\end{minipage}
	\label{fig:ContraSRates2}
    }
    	\hspace{3mm} 
    	\subfigure[]{
		\begin{minipage}[h]{0.3\linewidth}
			\includegraphics[width=5.7cm, height=3.2cm]{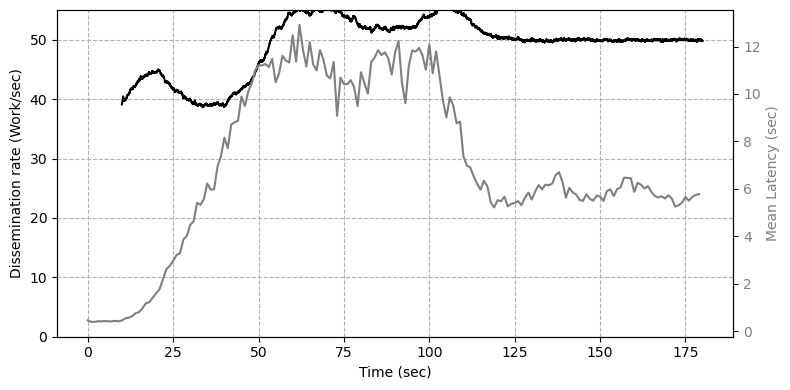}
		\end{minipage}
		\label{fig:Throughput9}
	}
	  	\hspace{1mm} 
    \subfigure[]{
    	\begin{minipage}[h]{0.3\linewidth}
   		\includegraphics[width=5.7cm, height=3.2cm]{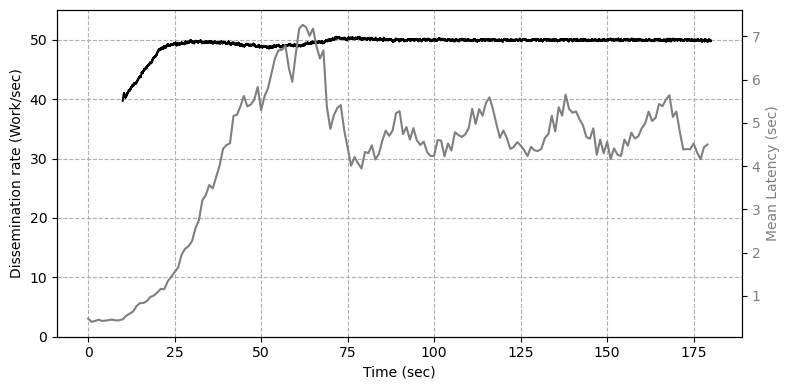}
    	\end{minipage}
	\label{fig:Throughput3}
    }
	\caption{(a) and (b) are dissemination rate and scaled dissemination rate in scenario which active nodes becoming inactive and reverting active, while the rest plots are contrast experiments with~\cite{cullenaccess}. (c) and (d) are dissemination rate and scaled dissemination rate.  (e) is dissemination rate and mean latency for each node in~\cite{cullenaccess}, while (f) is dissemination rate and mean latency for each node in this work. }
	\label{fig:Multiple_spamming}
\end{figure*}

Again, we retain the same network topology and reputation distribution of each node as in scenario A.(3). The dissemination rate of the network, and the mean latency for each node is depicted in Figure \ref{fig:Throughput9}. Comparing with Figure \ref{fig:Throughput3}, which depicts the situation when we have timestamp ordering, it is clear that the protocol is not robust without the timestamp ordering. This is much clearer from 50 seconds to 120 seconds. During this period of time, tn Figure \ref{fig:Throughput9}, the dissemination rate of all transactions is far over the maximum value 50 and the mean latency is also very high during this long time. In Figure \ref{fig:Throughput3}, while there is a slight increasing of mean latency, it converges to a stable value very quickly.
%
\section{Conclusions}
In this paper, an improved access control algorithm for DAG based DLT is designed to improve the security and robustness of such a network. A blacklisting algorithm, which is based on a reputation-weighted threshold, is introduced to handle both spamming and multi-rate malicious attackers. The introduction of a solidification request component is also introduced to ensure the fairness and consistency of network in the presence of attacks. Finally, a timestamp component is also introduced to maintain the consistency of the network in the presence of  multi-rate attackers. Simulations to illustrate the robustness of the protocol are also described. Future work will focus on maintaining network utilization in the presence of attacks.

\bibliographystyle{unsrt}
\bibliography{refs}

\end{document}